\documentclass[10pt, letterpaper]{article}

\usepackage{graphicx}
\usepackage{setspace}
\usepackage{tocloft}
\usepackage{float}
\usepackage{amsmath}
\usepackage{amsfonts}
\usepackage{array}
\usepackage{mathtools}
\usepackage{psfrag}
\usepackage[right]{lineno}
\usepackage[caption=false,font=footnotesize]{subfig}
\usepackage{epsfig}
\usepackage{threeparttable}
\usepackage{url}

\usepackage{lastpage,fancyhdr,graphicx}
\usepackage{epstopdf}
\fancyheadoffset[L]{2.25in}
\fancyfootoffset[L]{2.25in}

\graphicspath{{./}}

\DeclareMathOperator*{\argmin}{arg\,min}
\DeclareMathOperator{\sinc}{sinc}
\DeclareMathOperator{\supp}{supp}
\DeclareMathOperator{\sing}{sing}
\newtheorem{theorem}{\textbf{Theorem}}

\newtheorem{assertion}{\textbf{Assertion}}

\newcommand	\TStrut{\rule{0pt}{2.6ex}}       
\newcommand	\BStrut{\rule[-1.2ex]{0pt}{0pt}} 

\begin{document}
\vspace*{0.35in}

\begin{flushleft}
{\Large
\textbf\newline{Impact of non-stationary optical illumination on image reconstruction in optoacoustic tomography}
}
\newline
\\
Yang~Lou\textsuperscript{1},
Kun~Wang\textsuperscript{1},
Alexander~A.~Oraevsky\textsuperscript{2},
Mark~A.~Anastasio\textsuperscript{1, *},
\\
\bigskip
\bf{1} Department of Biomedical Engineering, Washington University in St. Louis, St. Louis, 63130
\\
\bf{2} TomoWave Laborotories, Houston, 77081
\\
\bigskip
* anastasio@wustl.edu

\end{flushleft}

\section*{Abstract}
	Optoacoustic tomography (OAT), also known as photoacoustic  tomography, is a rapidly emerging hybrid imaging technique that possesses great potential for a wide range of biomedical imaging applications. 
In OAT, a laser is employed to illuminate the tissue of interest and acoustic signals are produced via the photoacoustic effect.  
From these data, an estimate of the distribution of the absorbed optical  energy density within the tissue is reconstructed, referred to as the object function.  This quantity is defined, in part, by the distribution of light fluence within the tissue that is established by the laser source.
When performing three-dimensional imaging of large objects, such as a female human breast, it can be difficult to achieve a relatively uniform coverage of light fluence within the volume of interest when the position of the laser source is fixed.
To circumvent this, researchers have proposed illumination schemes in which the relative position of the laser source and ultrasound probe is fixed, and both are rotated together to acquire a tomographic data set.
A problem with this rotating-illumination scheme is that the tomographic data are  inconsistent; namely, the acoustic data recorded at each tomographic view angle (i.e., probe position) are produced by a distinct object function.   In this work, the impact of this data inconsistency on image reconstruction accuracy is investigated systematically.  This is accomplished by use of computer-simulation studies and application of mathematical results from the theory of microlocal analysis.  
These studies specify the set of image discontinuities that can be stably reconstructed with a non-stationary optical illumination set-up. 
The study also includes a comparison of the ability of iterative and analytic image reconstruction methods to mitigate artifacts attributable to the data inconsistency.

\section{Introduction}
\label{sec:intro}  

Optoacoustic computed tomography (OAT), also known as photoacoustic computed tomography (PACT), is a rapidly emerging hybrid imaging technique that has received wide-spread attention in the past decade \cite{PA_Origin_1, PA_Origin_2, Kun_2011_PAPrinciple, AlexanderBook, WangBook, WangReview, wang_basic_paper}. 
In OAT, biological tissues are irradiated by short laser pulses and generate internal acoustic wave fields via the photoacoustic effect. 
The propagated acoustic wave fields are detected by an array of ultrasound transducers surrounding the object. 
From the collected acoustic data, an OAT reconstruction method is employed to estimate the absorbed optical energy density, referred to as the \emph{object function}, within the tissue. 
Due to its hybrid nature, OAT achieves the high optical contrast of a pure optical imaging method with the high spatial resolution of a pure ultrasound imaging method. These advantages makes OAT highly desirable for biomedical imaging applications, among which breast cancer imaging is an important example.

The object function in OAT is determined by the product of  the light fluence distribution and optical absorption coefficient distribution within the object.
When performing three-dimensional (3D) imaging of  large objects, such as a female human breast, it is difficult to achieve a relatively uniform distribution of light fluence within the volume of interest when the position of the laser source is fixed \cite{illuminationSNR, JiangHuabei}.
To mitigate this problem,
  rotating illumination OAT (RI-OAT) system designs \cite{ermilov2016three,PeterMouse,SergeyMouse_2014,tsyboulski2014enabling, jetzfellner2010optoacoustic, Feng_RI_OAT} have been developed, which are the subject of investigation in this study. 
In the RI-OAT system design, the relative position of the laser fiber bundle and ultrasound probe is fixed, and both are rotated together around about a  scanning axis to acquire a tomographic data set.
Tomographic data recorded in this way are generally  inconsistent; namely, the acoustic data recorded at each tomographic view angle (i.e., probe position) are produced by a distinct object function because the light fluence distribution inside the object varies with view angle.  As described below, this presents challenges for image reconstruction.

When imaging relative small objects, the measurement data inconsistency in RI-OAT has not prevented informative imaging.  For example,  whole body small animal imaging using RI-OAT has been successfully employed to reveal complicated vascular and organ anatomy \cite{ermilov2016three,PeterMouse,SergeyMouse_2014,su2012optoacoustic} and distributions of molecular imaging probes \cite{tsyboulski2013towards, liopo2012optoacoustic,su2011gold}.   
A possible explanation for this is that, for sufficiently small cylindrically shaped objects that are oriented parallel to the axis of tomographic scanning, the light fluence distributions at different tomographic views do not differ greatly and the entire object is illuminated at each view, thereby reducing the degree of data inconsistency.  However, as shown below, this is not the case when imaging larger objects, in which case significant artifacts can be produced in RI-OAT.

A possible approach to eliminating the above mentioned data inconsistency in RI-OAT is to reformulate the reconstruction problem so that the optical absorption coefficient, rather than the absorbed optical energy density, is the to-be-reconstructed quantity (Guillaume Bal, personal communication, 2015, \cite{Feng_RI_OAT}).  
The inconsistency between the imaging model and measurement data would be removed because the optical absorption coefficient is an intrinsic property of the object that does not depend on the light distribution or tomographic view angle. 
This could be interpreted as a variant of the so-called quantitative OAT (Q-OAT) problem \cite{cox2006two,saratoon2013gradient,zemp2010quantitative,bal2011multi}. 
Jetzfellner et. al \cite{jetzfellner2010optoacoustic} and Feng \cite{Feng_RI_OAT} proposed reconstruction algorithms that sought to estimate the optical absorption coefficients of the object. 
Jetzfellner employed a simplified optical model that ignored optical heterogeneity within the object, whose validity may be compromised if a complicated object is to be considered; Feng assumed the diffusion approximation in the optical process and solved a joint optimization problem, but the proposed method is computationally intensive, especially when applied to real-world 3D problems. 
Due to the limitations of currently available RI-OAT reconstruction methods, most current implementations of RI-OAT still employ conventional OAT reconstruction methods that assume stationary light illumination \cite{xu2005universal,wang2011imaging,KunMouse,wang2014discrete}.
While a recent mathematical analysis has been reported by Bal and Moradifam \cite{BalDraft}, there have been no reported numerical investigations that demonstrate the limitations of  conventional OAT image reconstruction methods when employed in RI-OAT.   
There remains a significant need for such an investigation, as it would  reveal the effectiveness of RI-OAT for clinical imaging applications and could affect future system designs.




 In this work, the impact of  data inconsistency in RI-OAT on image reconstruction accuracy is investigated. 
 This is accomplished by use of computer-simulation studies and application of mathematical results from the theory of microlocal analysis.  
 Sufficient conditions for stable reconstruction of singularities (i.e., edges) in the object function are identified.
 A study that compares the ability of iterative and analytic image reconstruction methods to mitigate artifacts attributable to the data inconsistency is also presented.

The paper is organized as follows. In Section \ref{sec:background}, we review the canonical OAT imaging model  and salient mathematical results from the theory of microlocal analysis.
Theoretical insights into the RI-OAT reconstruction problem are provided in Section \ref{sec:theoryCC}. 
These include a statement and interpretation of the RI-OAT imaging model in its continuous form and application of microlocal analysis concepts to identify which image boundaries can be stably reconstructed.
Section \ref{sec:method} describes the computer-simulation studies that were designed to systematically corroborate the theoretical insights, with the numerical results being presented in Section \ref{sec:simulation}.
Finally,  a discussion of the study and conclusions are provided in Section  \ref{sec:discussion}.

\section{Background}
\label{sec:background}
Below, the canonical  OAT imaging model in its continuous and discrete forms along with a motivation for employing RI-OAT are reviewed.
Salient results from the theory of microlocal analysis  \cite{sonar, Quinto_PAT, Linh_PAT, Quinto_LimitAngle, nguyen2014reconstruction} are also presented. 

\subsection{Canonical OAT imaging models in continuous and discrete forms}
\label{subsec:imaging_model}
In OAT, a pulsed laser source irradiates the object. If the optical pulse duration is short compared to the thermal relaxation time of the material, 
absorption of the optical energy will produce an acoustic pressure field via the photoacoustic effect \cite{PA_Origin_1, PA_Origin_2, AlexanderBook, WangReview, wang_basic_paper, WangBook}.
This pressure field will be denoted by the twice-differentiable function  $p(\mathbf{r}, t)$,
 where $\mathbf{r} \in \mathbb{R}^3$ denotes a 3D spatial coordinate, and $t\in \mathbb{R}_+$ is the temporal coordinate. 
The physical model of acoustic signal generation is given by the wave equation \cite{wang_basic_paper} 
\begin{equation}
	\left[ \frac{\partial ^2}{\partial t^2} - c^2 \bigtriangleup \right] p(\mathbf{r}, t) = 0, 
	\label{eq:wave_equation}
\end{equation}
subject to the initial conditions 
\begin{equation}
	p(\mathbf{r}, t) \vert_{t=0} = \frac{\beta c^2}{C_p} A(\mathbf{r}) = \Gamma A(\mathbf{r}), \qquad	\frac{\partial p(\mathbf{r}, t)}{\partial t} \bigr \vert_{t=0} = 0, 
	\label{eq:wave_condition}
\end{equation}
where the compactly supported and bounded function $A(\mathbf{r})$ represents the absorbed optical energy density,
also referred to as the object function, $c$ denotes the speed of sound, and $\bigtriangleup$ denotes the 3D Laplacian operator. 
The mathematical model that describes the relationship between $A(\mathbf{r})$ and optical properties of the object along with incident light field is detailed in Appendix \ref{append:light}.
In Eqn. (\ref{eq:wave_condition}), $\beta$ denotes the thermal expansion coefficient, $C_p$ is the heat capacity at a constant pressure, and $\Gamma$ is the dimensionless Gruneisen parameter. 
In our study, the acoustic properties within the object and background medium, including speed of sound and density, are assumed to be homogeneous within the object.
Therefore the quantities $c$, $\beta$, $C_p$, and $\Gamma$ can be regarded as constants independent of location $\mathbf{r}$. 

The sought-after object function can be expressed as $A(\mathbf{r})=\Phi(\mathbf{r})\mu_a(\mathbf{r})$, where $\Phi(\mathbf{r})$ and $\mu_a(\mathbf{r})$ denote the distribution of light fluence and optical absorption coefficient, respectively, within the object.
 As described in Sec.\ \ref{sec:theoryCC}, image reconstruction in RI-OAT is hampered by the fact that $\Phi(\mathbf r)$, and hence $A(\mathbf r)$, varies as a function of tomographic view angle.

 \subsubsection{\textbf{Continuous-to-continuous (C-C) imaging model}}
 Assuming point-like idealized ultrasonic transducers are employed,  the solution of Equation.\ (\ref{eq:wave_equation}) with initial conditions in (\ref{eq:wave_condition})
specifies the recorded pressure signal  at transducer location $\mathbf{r^\prime}$ as  \cite{WangBook, xu2005universal, finch2004determining}
\begin{equation}
	p(\mathbf{r}', t) = R_S(A(\mathbf{r}))\equiv
																		 \frac{\beta}{4 \pi C_p} \int_V \mathrm{d} \mathbf{r} A(\mathbf{r}) \frac{\partial}{\partial t} \frac{\delta(t - \frac{| \mathbf{r' - r} |}{c})}{|\mathbf{r' - r}|},    \label{eq:ContModel_2}
\end{equation}
where $V$ is the support of the object. 
Here, $R_S$ denotes the OAT forward operator and $\delta(t)$ is the one-dimensional Dirac-delta function. 
Equation \ (\ref{eq:ContModel_2}) represents a canonical continuous-to-continuous (C-C) imaging model for OAT. 

\subsubsection{\textbf{Continuous-to-discrete (C-D) imaging model}}
In practice, the ultrasonic transducer employed to record  $p(\mathbf{r}', t)$ has a finite-sized detection area and imperfect temporal response \cite{wang2011imaging, rosenthal2011model, cox2007frequency}. 
 The transducer surface in this study is assumed to be flat and samples of $p(\mathbf{r}', t)$ are recorded with a  sampling interval of $\Delta T$.
Consider that $Q$ transducers are positioned at locations $\{ \mathbf{r}_q, q = 0, 1, 2, \dots, Q-1 \}$ and $K$ temporal samples are recorded at each.
The $Q\times K$-dimensional vector $\mathbf{u}$ will denote the complete set of measurements that have been lexicographically ordered.
The $k$-th  temporal sample recorded by the $q$-th transducer is described as $[\mathbf{u}]_{qK + k}$.

As with any digital imaging system,
 a continuous-to-discrete (C-D) imaging model fundamentally describes the data-acquisition process in  OAT.
A C-D imaging model for OAT
 that maps the object function to the collection of measured data samples
can be  expressed as \cite{Kun_2011_PAPrinciple, wang2014discrete, KunGPU} 
\begin{equation}
	[\mathbf{u}]_{qK + k} = h^e(t) *_t \frac{1}{\Omega_q} \int _{\Omega_q} \mathrm{d} \mathbf{r'} p(\mathbf{r'}, t) \Bigr \vert_{t = k \Delta T} ,
	\label{eq:u_def_time}
\end{equation}
where $p(\mathbf{r'}, t)$ is determined by the object function $A(\mathbf r)$ via (\ref{eq:ContModel_2}),
 $h^e (t)$ denotes the electrical impulse response (EIR) of the transducer, $*_t$ denotes a temporal convolution, and $\Omega_q$ denotes the surface of the $q$-th transducer element. 

 \subsubsection{\textbf{Discrete-to-discrete (D-D) imaging model}}
To obtain a fully discretized imaging model  for use with iterative image reconstruction methods,
the object function $A(\mathbf{r})$ can be approximated by use of a finite  collection of expansion functions ${\phi_n}$ as \cite{wang2014discrete}
\begin{equation}
        A(\mathbf r)\approx A^a(\mathbf{r}) = \sum_{n = 0}^{N-1} [\boldsymbol{\theta}]_n \phi_n(\mathbf{r}),
\end{equation}
where $ [\boldsymbol{\theta}]_n$ is the $n$-th element of the $N$-dimensional coefficient vector $\boldsymbol{\theta}$.
In the numerical studies below,  uniform spherical voxel expansion functions are employed that are defined as \cite{lewitt1990multidimensional, schweiger2003image}
\begin{equation}
	\phi_n(\mathbf{r}) = \left \{ \begin{array}{cl}
			1, &\| \mathbf{r} - \mathbf{r}_n \| \leq \epsilon ,		\\
			0, &\textrm{otherwise. }
	\end{array} \right .
	\label{eq:expansion}
\end{equation}
In order to incorporate the directivity of the finite-size transducer element, the transducer's spatial impulse response (SIR) can be incorporated in the forward model.   To accomplish this, it will be convenient to consider the discrete Fourier transform of the measurement data $\mathbf u$, which will be denoted by the vector $\mathbf {\tilde{u}}$.  More specifically, let  $u_q(t)$ denote the pre-sampled voltage signal corresponding to the $q$-th transducer and let  $\tilde{u}_q(f)$ denote its temporal Fourier transform.   Consider that  $L$ samples of $\tilde{u}_q(f)$ are acquired with a sampling interval of $\Delta f$.
The $Q\times L $ dimensional data vector $\mathbf{\tilde{u}}$ represents a lexicographically ordered representation of the sampled frequency data, corresponding to all transducer locations, , i.e., $[\tilde{\mathbf{u}}]_{qL + l} \coloneqq \tilde{u}_q(f) \bigr \vert _{f = l\Delta f} $.

In terms of the quantities introduced above,
a discrete-to-discrete (D-D) imaging model can be expressed as 
\begin{equation}
	\mathbf{\tilde{u}} = \mathbf{H} \boldsymbol{\theta},
	\label{eq:discrete_basic}
\end{equation}
where $\mathbf{H} \in \mathbb{R}^{M \times N}$ is the system matrix with $M = Q\times L$ denoting the total number of temporal frequency samples.
 The elements of $\mathbf{H}$ are given by \cite{KenjiPaper}: 
\begin{gather}
	[\mathbf{H}]_{qL+l, n} = p_0(f) \tilde{h}^e(f) \frac{h_q^s(\mathbf{r}_n, f)}{ab} \Bigr \vert_{f=l\Delta f}	 	,  \label{eq:system_matrix} \\
	\textrm{with} \quad p_0(f) = -\mathrm{i} \frac{\Gamma c}{f}  \left[ \frac{\epsilon}{c} \cos \frac{2 \pi f \epsilon}{c} - \frac{1}{2 \pi f} \sin \frac{2 \pi f \epsilon}{c} \right]		, \\
	\quad  h_q^s(\mathbf{r}_n, f) = ab \frac{e^{-\mathrm{i} 2\pi f r_{n,q} /2}}{2\pi r_{n, q}} \sinc \bigl( \pi f \frac{a X_{n,q}}{c r_{n,q}} \bigr) \sinc \bigl ( \pi f \frac{b Y_{n,q}}{c r_{n,q}} \bigr ) .		\label{eq:SIR}
\end{gather}
Here, $p_0(f)$ is the temporal Fourier transform of the pressure data produced by a spherical voxel of radius $\epsilon$, $\tilde{h}^e(f)$ is the temporal Fourier transform of the EIR $h^e(t)$, and $\mathrm{i}\equiv\sqrt{-1}$.  The quantity
 $h_q^s(\mathbf{r}_n, f)$ in (\ref{eq:SIR}) describes the temporal Fourier transform of the SIR of the $q$-th transducer under a far-field approximation \cite{wang2011imaging, KenjiPaper}, where
$\mathbf{r}_n$ is the position of the $n$-th voxel,  and $a$ and $b$ are the dimensions of the planar transducer element. 
The distance between the $n$-th voxel and the center of the $q$-th transducer is denoted by $r_{n,q} = \| \mathbf{r}_n - \mathbf{r'}_q \|$. 
The quantities $X_{n,q}, Y_{n,q}$ are the coordinates of the $n$-th voxel in the local coordinate system centered at the $q$-th transducer position \cite{KenjiPaper}.


\begin{figure}[htbp]
        \centering
        \includegraphics[width=0.8\linewidth]{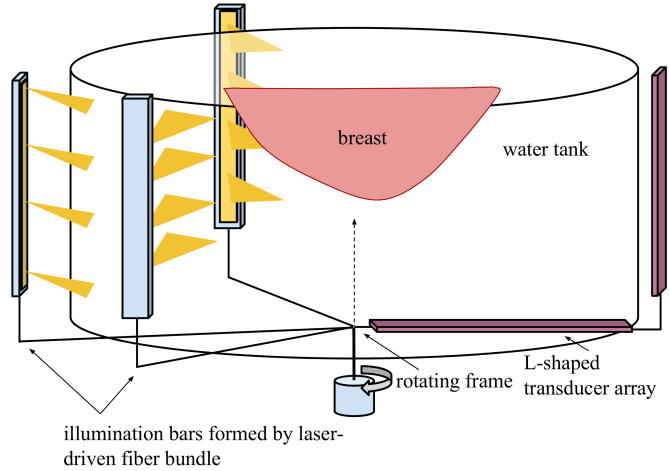}
        \caption{A schematic of a 3D breast imaging system employing a rotating partial illumination design.}
        \label{fig:system}
\end{figure}

\subsection{Motivation for RI-OAT}

An example of a possible RI-OAT system design for breast imaging is depicted in Fig.\ \ref{fig:system}.
A similar RI design has been successfully implemented for small animal imaging \cite{SergeyMouse_2014, PeterMouse, KunMouse}. 
The breast is immersed in an imaging module that is filled with water.  The imaging module contains a light delivery system and an acoustic probe.
To deliver light, laser pulses are directed into optical fibers, which are bundled into rectangular illumination bars that redirect the light  toward the breast.  The acoustic probe is located on the opposite side of the breast, which is employed to record the induced optoacoustic signals.  During tomographic scanning, the light delivery system and acoustic probe are simultaneously rotated about the breast over a full 360$^\circ$ angular range. 
As such, the light fluence $\Phi(\mathbf r)$ varies with tomographic view angle. This indicates that
the system is, in effect, recording pressure data produced by different object functions at each view angle.

From an implementation perspective, however,
this design is not without merit.
By delivering the light into a select region of the breast instead of attempting to illuminate the entire breast, for a given laser power, more light can penetrate to deeper locations within the breast. 
In particular, regions of the breast  near the chest wall would be better illuminated \cite{Yang_signal_detection}.
The design could be modified in such a way that, for a given acoustic probe location, the illumination system is rotated around the breast to produce a more uniform illumination pattern. However, this would increase scan times.
The purpose of the investigations reported below is to better understand the limitations of RI-OAT and identify what type of information can be reliably obtained when a conventional OAT reconstruction method is employed.


\subsection{Relevant results from  the theory of microlocal analysis as applied to OAT} 
\label{subsec:microlocal}

As will be demonstrated in Sec.\ \ref{sec:theoryCC}, when a conventional OAT image reconstruction method is employed in RI-OAT, the reconstructed image will not, in general, represent an accurate estimate of $A(\mathbf r)$. However, the reconstructed image can contain
reliable information regarding the locations of sharp edges or discontinuities in $A(\mathbf r)$; this information can facilitate a variety of diagnostic tasks related to the detection and characterization of anatomical structures. 
In a mathematical sense, such features can be interpreted as singularities in $A(\mathbf r)$ \cite{RaamBook,Faridani:01}.
The \emph{wavefront set} of a function, which is a central concept
in the theory of microlocal analysis \cite{MicroBook,Quinto:93},
can comprehensively characterize the singularities in a function.
When a C-C imaging model is assumed (e.g., Eqn.\ (\ref{eq:ContModel_2})), results from 
the theory of microlocal analysis can be employed to identify the subset of the wavefront set of
 $A(\mathbf r)$ that can be stably reconstructed from a given set of OAT measurements 
\cite{pan2003data,xu2004reconstructions,frikel2015artifacts, Quinto_PAT, nguyen2014reconstruction, nguyen2015artifacts, Linh_PAT, kuchment2011mathematics,sonar}.

\subsubsection{\textbf{The wavefront set}}
The regularity, or smoothness, properties of a function are reflected in the decay properties of its Fourier transform \cite{Faridani:01}. 
Specifically, the more smooth a function is, as measured by the existence of its partial derivatives, the more rapidly its Fourier transform will decay.
Functions that contain discontinuities will possess Fourier transforms that decay less rapidly than those corresponding
to functions that do not.
This indicates that the {global} smoothness properties of a
function can be inferred by examination of its Fourier
transform decay properties \cite{RaamBook}.

Characterization of singularities can be accomplished by 
extending these concepts
to examine a localized region of a function
in the following manner.
Let $\phi(\mathbf {r})$ denote a compactly supported and infinitely differentiable
window function that satisfies $\phi(\mathbf {r}_s)\ne 0$ and is
zero outside of some neighborhood of $\mathbf r_s$.
The function  $A_\phi(\mathbf r)\equiv\phi(\mathbf r)A(\mathbf r)$ will have the same
singularities (if there are any) as $A(\mathbf {r})$ near $\mathbf {r}_s$,
and will equal zero away from $\mathbf {r}_s$.
Let $\tilde{A}_\phi(\boldsymbol {\nu})$ denote the 3DFourier transform of  $A_\phi(\mathbf {r})$.
If $A(\mathbf {r})$, and hence $A_\phi(\mathbf {r})$, possesses
a singularity at $\mathbf {r}_s$, it will be reflected in the
decay properties of $\tilde{A}_\phi(\boldsymbol{\nu})$.
Specifically,  if $\tilde{A}_\phi(\boldsymbol{\nu})$
does not decay sufficiently rapidly, in {\it all} directions, we know that a singularity
exists in $A(\mathbf {r})$ at or near location $\mathbf {r}_s$.

The direction of a singularity at location $\vec{\mathbf{r}}_s$ in $A(\mathbf {r})$ is defined as the direction in which $\tilde{A}_\phi(\boldsymbol {\nu})$ does not decay sufficiently rapidly.   
Therefore, a singularity can be described completely by its spatial location $\mathbf {r}_s$ and
the direction in which the Fourier transform of the associated localized function does not decay sufficiently rapidly, which we denote by $\xi(\mathbf r_s)$.
The wavefront set of $A(\vec{r})$, denoted by $WF(A)$,
is defined by the elements $\{(\mathbf {r}_s,\xi(\mathbf {r}_s))\}$, and
provides a complete characterization of the singularities
in $A(\vec{r})$.
A more formal mathematical definition of the wavefront set is provided in Appendix \ref{append:wave_front_set}.

\subsubsection{\textbf{A microlocal correspondence for OAT}}
Microlocal correspondences follow from fundamental results in microlocal analysis and
 provide valuable insights into the object features that can be stably reconstructed in a tomographic inverse problem.
More precisely, microlocal correspondences provide a relationship between the wavefront set of an object function and
the wavefront set of the tomographic data function.  For OAT, the microlocal correspondence yields
a simple geometric interpretation that is given as follows \cite{sonar, frikel2015artifacts, kuchment2011mathematics, Anastasio_truncation}.

\vspace{2ex}
	\begin{theorem}
	\emph{A wave front set component $(\mathbf{r},\xi)$ of $A(\mathbf{r})$ is stably recoverable (or detectable) from the pressure data $R_S [A(\mathbf{r})]$ measured on a continuous measurement surface $S$ if and only if the line extended in both directions by $\xi$ intersects the interior of $S$.}
	\end{theorem}
\vspace{2ex}

\begin{figure}[htbp]
	\centering
	\includegraphics[width = 0.6\linewidth]{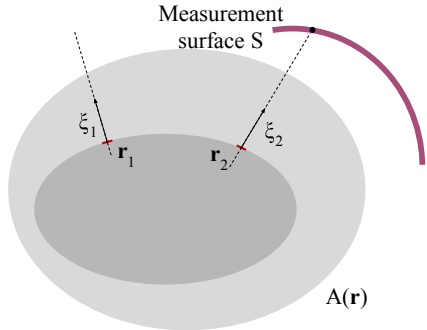}
	\caption{An illustration of the concept of wave front set and how it relates to Theorem 1. The vectors $\mathbf{r}_1$
 and $\mathbf{r}_2$ denote points on the boundary of $A(\mathbf{r})$. The vectors
$\xi_1$ and $\xi_2$ are normal to the boundary at $\mathbf{r}_1$ 
 and $\mathbf{r}_2$.  The measurement aperture is denoted by  the surface $S$. }
	\label{fig:wave_front_set}
\end{figure}

This theorem states that a singularity of the object function can be \emph{stably detected} if and only if the wave propagated from that singular point along its normal can be detected by the measurement surface. 
A simple illustration of this concept is provided in Fig. \ref{fig:wave_front_set}. Here, the object function $A(\mathbf{r})$ is piecewise constant. Both $(\mathbf{r}_1, \xi_1)$ and $(\mathbf{r}_2, \xi_2)$ are examples of wave front components of $A(\mathbf{r})$. According to Theorem 1, only $(\mathbf{r}_2, \xi_2)$ is stably recoverable because the line extended by $\xi_2$ intersects with measurement surface $S$, while the line extended by $\xi_1$ does not.

\section{RI-OAT imaging model and interpretation: Continuous case}
\label{sec:theoryCC}
Below, the RI-OAT imaging model in its continuous form is introduced, along with an interpretation of the image reconstructed by use of a conventional reconstruction method that assumes a fixed optical illumination. 
The theory of microlocal analysis is applied to RI-OAT to determine the stably recoverable singularities within the object under both different illumination conditions. 

\subsection{C-C imaging model for RI-OAT}
To present the C-C imaging model for RI-OAT, we consider the three dimensional (3D) case. 
The acoustic probe contains a single transducer element at position $\mathbf{r}_0$, and the measurement data are collected by moving the probe on a measurement surface $S_0$. 
Without loss of generality, it will also be assumed that the measurement surface $S_0$ is a sphere that encloses the to-be-imaged object. 
As described above, in RI-OAT, the light fluence distribution and consequently the object function are both functions of the acoustic probe location $\mathbf{r}_0$.
Accordingly, a view-dependent object function $A(\mathbf{r}; \mathbf{r}_0)$ can be defined as: 
\begin{equation}
	A(\mathbf{r}; \mathbf{r}_0) = \Phi(\mathbf{r}; \mathbf{r}_0) \mu_a(\mathbf{r}), 
	\label{eq:A_r_RI_1}
\end{equation}
where $\mu_a(\mathbf{r})$ is the optical absorption coefficient and $\Phi(\mathbf{r}; \mathbf{r}_0)$ is the view-dependent light fluence distribution. 
The corresponding pressure data at measurement location $\mathbf{r}_0 \in S_0$, denoted by $p^{RI}(\mathbf{r}_0, t)$, are given by: 
\begin{equation}
	p^{RI}(\mathbf{r}_0, t) = \frac{\beta}{4\pi C_p}\int_V \mathrm{d} \mathbf{r} \, A(\mathbf{r}; \mathbf{r}_0)h(\mathbf{r}, t; \mathbf{r}_0), 
	\label{eq:p_RI_1}
\end{equation}
where $h(\mathbf r, t;\mathbf r_0)\equiv \frac{\partial}{\partial t} \frac{\delta(t - \frac{| \mathbf{r_0 - r} |}{c})}{|\mathbf{r_0 - r}|}$. 

In practice, measurement data are acquired at a finite number of tomographic views. 
Let $J$ denote the total number of views, and let the measurement surface $S_0$ be evenly divided into $J$ surface elements: $\{S_j, j = 1, 2, \dots, J \}$.
The $j$-th surface element $S_j$ is the section of the sphere whose polar angular range is $[\phi_j, \phi_{j+1}]$, where $\phi_j = j \cdot \frac{2 \pi}{J}$, and whose azimuth angular range is $[0, \pi]$. 
If the polar angular span for each surface element is small, or equivalently, the view number $J$ is large, the light fluence distribution, and thus the object function, is approximately constant when the acoustic probe is within each surface element $S_j$, which motivates the following assumption: 
\begin{description}
	\item [\textbf{Assumption 1.}] When $\mathbf{r}_0 \in S_j$, the light fluence distribution is fixed and independent of $\mathbf{r}_0$ and is denoted by $\Phi_j(\mathbf{r})$. 
		The corresponding absorbed optical energy is also independent of $\mathbf{r}_0$ and is denoted by $A_j(\mathbf{r}) = \Phi_j(\mathbf{r}) \mu_a(\mathbf{r})$. 
\end{description}

The corresponding recorded pressure data are given by:  
\begin{gather}
	p^{RI}(\mathbf{r}_0, t) \approx \frac{\beta}{4 \pi C_p} \int_V \mathrm{d} \mathbf{r}\, A_j(\mathbf{r}) h(\mathbf r, t;\mathbf r_0), \text{ when } \mathbf{r}_0 \in S_j.			\label{eq:p_RI_2}
\end{gather}
Assumption 1 and Equation (\ref{eq:p_RI_2}) indicate that the measurement data recorded within the $j$-th surface element $S_j$ are produced by the $j$-th object function $A_j(\mathbf{r})$. 
Hence the forward imaging process and the reconstruction problem of RI-OAT can be regarded as a combination of $J$ limited-angle tomography subproblems, each involving a view-specific object function $A_j(\mathbf{r})$.

\subsection{Interpretation of the reconstructed image}
\label{subsec:interpretation}
Although Eqn. (\ref{eq:p_RI_2}) suggest that the RI-OAT reconstruction problem can be formulated into $J$ limited-view subproblems, each subproblem suffers from severe data-incompleteness, hence there are currently no reconstruction methods that can effectively invert Eqn. (\ref{eq:p_RI_2}) and estimate $A_j(\mathbf{r})$ for all $j$. 
Most reported experimental implementations of RI-OAT have employed standard OAT reconstruction methods that ignore the fact that the light fluence distribution changes with tomographic view angle. 
Below we interpret the reconstructed image when a filtered back-projection (FBP) operator is employed. 

The image reconstructed by application of a FBP inversion formula to the measurement data $p^{RI}(\mathbf r_0,t)$ can be expressed as
\begin{equation}
	\hat{A}^{RI}(\mathbf r)\equiv \eta \int_{S_0} dS_0\; T\left[ p^{RI}(\mathbf r_0,t)\right ]_{t=\frac{\vert\mathbf r -\mathbf r_0\vert}{c}}, 
	\label{eq:RIrecon}
\end{equation}
where $\mathbf r_0 \in S_0$, $dS_0$ is a solid angle differential, and $\eta$ is a constant. 
Here, $T$ is an appropriately defined filtering operator acting on the temporal coordinate of the pressure data function so that Eqn. (\ref{eq:RIrecon}) represents an exact inversion formula  \cite{xu2005universal}. 
The explicit form of Eqn. (\ref{eq:RIrecon}) is not important; the analysis that follows is valid for any choice of a mathematically exact inversion formula in the form of a FBP operator.

For each limited-view subproblem specified by Eqn. (\ref{eq:p_RI_2}), application of the FBP formula to the limited-view measurement data recorded within $S_j$ yields
\begin{equation}
	\hat{A}_j(\mathbf{r}) = \eta \int_{S_j} dS_0 T[p^{RI}(\mathbf{r}_0, t)]_{t = \frac{|\mathbf{r} - \mathbf{r}_0|}{c}} ,
	\label{eq:partial}
\end{equation}
where $\hat{A}_j(\mathbf{r})$ denotes a partial reconstructed image corresponding to the $j$-th limited-view subproblem. 
By use of Eqns. (\ref{eq:RIrecon}) and (\ref{eq:partial}), it is readily verified that
\begin{equation}
	\hat{A}^{RI} (\mathbf{r}) = \sum_{j=1}^{J} \hat{A}_j(\mathbf{r}).
	\label{eq:RI_summation}
\end{equation}
This establishes that the image reconstructed in RI-OAT by use of a conventional linear reconstruction method can be interpreted as superposition of estimates of $A(\mathbf r)$ that are reconstructed from each of the $J$ limited-view subproblems.
A fully discrete version of the analysis presented above is provided in Appendix B. 

\subsection{Stable detection of singularities with sufficient illumination}
\label{subsec:stable}
Because the object function $A_j(\mathbf{r})$ varies for different tomographic view $j$, it is not useful to directly analyze its wavefront set.
However, noting that $\mu_a(\mathbf{r})$ is an intrinsic property of the object and is independent of the view angle, the wavefront set of $\mu_a(\mathbf{r})$ instead of $A_j(\mathbf{r})$ is considered below. 
We establish a relation between the singularities in $\mu_a(\mathbf{r})$ and singularities in the estimate $\hat{A}^{RI}(\mathbf{r})$ obtained by use of the FBP inversion formula under the following assumptions:  
\begin{description}
	\item [\textbf{Assumption 2.}] The fluence distribution $\Phi_{j}(\mathbf{r})$ is smooth with respect to $\mathbf{r}$, for $j = 1, 2, \dots, J$, which is equivalent to saying $WF[\Phi_j(\mathbf{r})] = \emptyset$. 
	\item [\textbf{Assumption 3.}] The fluence distribution $\Phi_{j}(\mathbf{r}) > 0$, for $\mathbf{r} \in V$. 
\end{description}
Note that these assumptions hold true in a mathematical sense when $\Phi_j(\mathbf{r})$ is a solution of the radiative transfer equation (RTE) (\cite{WangBook}). 
In practice, these assumptions indicate a \textit{sufficient illumination} condition, where light can penetrate through the entire to-be-imaged object and generate non-zero optoacoustic signals within the object.  
The case where Assumption 3 does not hold is referred to as an insufficient illumination condition and will be addressed in section \ref{subsec:stable_practical}. 

Under these assumptions, one obtains 
\begin{equation}
	WF[\mu_a(\mathbf{r})] = WF[A_j(\mathbf{r})], 
	\label{eq:WF_equal}
\end{equation}
indicating that when $\Phi_j(\mathbf{r})$ is smooth and non-zero within $V$, all the singularities in $\mu_a(\mathbf{r})$ will be contained in the object function $A_j(\mathbf{r})$, although their magnitudes may differ. 

To proceed, the notions of `visible singularities' and `added singularities' will be employed \cite{Linh_PAT, Quinto_PAT}. 
For a wavefront component $(\mathbf{r}, \xi)$, let us denote the line that passes through $\mathbf{r}$ and parallel to $\xi$ as $l(\mathbf{r}, \xi)$. 
Denote the boundary of the measurement surface $S$ as $\partial S$, and the interior of $S$ as $S_{int} = S \backslash \partial S$. 
Denote the sphere centered at $\mathbf{r}_1$ and passes through $\mathbf{r}_2$ as $C(\mathbf{r}_1, \mathbf{r}_2)$. 

Consider the canonical OAT imaging process described in Eqn. (\ref{eq:ContModel_2}), with measurement data recorded on a smooth and convex measurement surface $S$. 
Consider an object function $A(\mathbf{r}):\mathbb{R}^3 \rightarrow \mathbb{R}$ that has a finite support $V$. 
		The set of \textbf{visible singularities} in $A(\mathbf{r})$ given $S$ is defined as \cite{Linh_PAT, Quinto_PAT}:
		\begin{equation}
			\mathcal{V}_S(A(\mathbf{r})) = \{ (\mathbf{r}, \xi) \in WF[A(\mathbf{r})], \: l(\mathbf{r}, \xi) \text{ intersects } S_{int} \},  			
			\label{eq: visible}
		\end{equation}
		These singularities correspond to the stably recoverable singularities defined in Theorem 1. 
		
If there exists a $(\mathbf{r}_A, \xi_A) \in WF[A(\mathbf{r})]$ such that $l(\mathbf{r}_A, \xi_A)$ intersects with $\partial S$, denote the point of intersection as $\mathbf{r}_S \in \partial{S}$. 
		Consider the sphere $C(\mathbf{r}_S, \mathbf{r}_A)$.		
		The set of \textbf{added singularities} is defined as \cite{Linh_PAT, Quinto_PAT}:
		\begin{gather}
			\mathcal{A}_S(A(\mathbf{r})) = \{ (\mathbf{r}, \xi):  \mathbf{r} \in C(\mathbf{r}_S, \mathbf{r}_A), 	\nonumber 	\\
			\text{and } \xi \text{ is the outward normal vector of } C(\mathbf{r}_S, \mathbf{r}_A) \text{ at } \mathbf{r}. \}
		\label{eq:added}
		\end{gather}
 results in microlocal theory, we have the following \cite{Quinto_PAT, Linh_PAT}: 

\begin{theorem}
	\emph{ 
		Denote the forward OAT imaging operator in Eqn. (\ref{eq:ContModel_2}) as $R_S$, with pressure data recorded on a smooth and convex measurement surface $S$. 
		Consider an object function $A(\mathbf{r}): \mathbb{R}^3 \rightarrow \mathbb{R}$ with a finite support $V$. 
		Let $B$ denote the filtered back-projection operator. 
		The singularities in the original function $A(\mathbf{r})$ and the reconstructed estimate $B R_S A(\mathbf{r})$ are related as:
	\begin{equation}
		\mathcal{V}_S(A(\mathbf{r})) \subset WF[B R_S A(\mathbf{r})] \subset \{ \mathcal{V}_S(A(\mathbf{r})) \cup \mathcal{A}_S(A(\mathbf{r})) \}.
		\label{eq:thm_1}
	\end{equation} }
\end{theorem}

\begin{figure}[htbp]
	\centering
	\includegraphics[width = 0.9\linewidth]{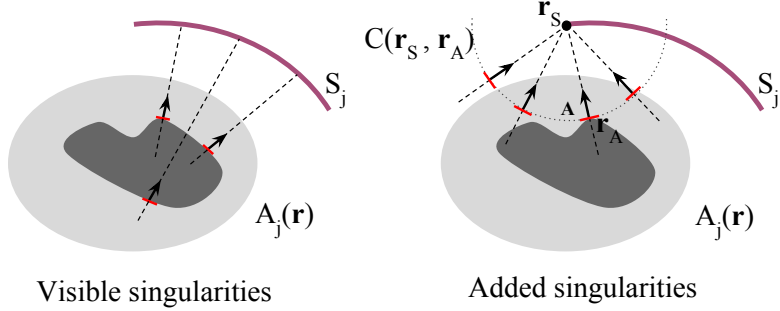}
	\caption{Illustration of visible singularities and added singularities. }
	\label{fig:microlocal_2}
\end{figure}

By substituting $A(\mathbf{r})$ with $A_j(\mathbf{r})$ and $S_j$ with $S$ in Eqn. (\ref{eq:thm_1}) one obtains:
\begin{equation}
	\mathcal{V}_{S_j}(A_j(\mathbf{r})) \subset WF[\hat{A}_j(\mathbf{r})] \subset \{ \mathcal{V}_{S_j}(A_j(\mathbf{r})) \cup \mathcal{A}_{S_j}(A_j(\mathbf{r}))\}. 
	\label{eq:A_j_thm}
\end{equation}
Equation (\ref{eq:A_j_thm}) states that for each aforementioned limited-angle subproblem in RI-OAT, the reconstructed estimate $\hat{A}_j(\mathbf{r})$ will contain all the visible singularities in $A_j(\mathbf{r})$, but will also possibly contain additional singularities that represent artifacts. 
An illustration of this relationship in 2D is given in Fig. \ref{fig:microlocal_2}. 
The visible singularities in $A_j(\mathbf{r})$ are the ones for which $l(\mathbf{r}, \xi)$ intersects the interior of $S_j$.
The added singularities can be intuitively explained as follows: 
if $l(\mathbf{r}, \xi)$ intersects the boundary of $S_j$ at $\mathbf{r}_{S}$, then when the FBP operator is `back-projecting' to form a reconstructed image, it not only back-projects the singularity to the original location at $\mathbf{r}$, but also back-projects to the entire sphere that is centered at $\mathbf{r}_{S}$ and passes through $\mathbf{r}$, thereby forming the added singularities. 

Equations. (\ref{eq:RI_summation}), (\ref{eq:WF_equal}), and (\ref{eq:A_j_thm}) establish that 
\begin{multline}
	\cup_{j=1}^{J} \mathcal{V}_{S_j}(\mu_a(\mathbf{r})) \subset WF[\hat{A}^{RI}(\mathbf{r})] \\ 
	\subset (\cup_{j=1}^{J} \mathcal{V}_{S_j}(\mu_a(\mathbf{r}))) \cup (\cup_{j=1}^{J} \mathcal{A}_{S_j}(\mu_a(\mathbf{r}))). 
	\label{eq:union_thm}
\end{multline}
This result yields the following assertion:

\vspace{2ex}
	\begin{assertion}
		\emph{
			In RI-OAT, under Assumptions 1, 2, and 3, a wave front set component $(\mathbf{r}, \xi)$ of $\mu_a(\mathbf{r})$ can be stably recovered by forming the estimate $\hat{A}^{RI}(\mathbf{r})$, if and only if the line extended in both directions by $\xi$ intersects the interior of one of the measurement surface elements $S_j$. Meanwhile, $\hat{A}^{RI}(\mathbf{r})$ may include additional singularities that represent artifacts, whose locations and directions are given by Eqn. (\ref{eq:added}) with $A(\mathbf{r}) = \mu_a(\mathbf{r})$. 
	}
	\end{assertion}
\vspace{2ex}

Note that in the special case of stationary (and sufficient) illumination, the microlocal correspondence that forms the basis for Assertion 1 reduces to that in Theorem 1. 
In that case, when $S_0$ is a smooth convex surface that completely encloses the object, all singularities in $\mu_a(\mathbf{r})$ are visible: $\mathcal{V}_{S_0}({\mu_a(\mathbf{r})}) = WF[\mu_a(\mathbf{r})]$, and will be recovered in the estimate $\hat{A}^{RI}(\mathbf{r})$.  
The added singularities for the $J$ subproblems are completely cancelled out when they are summed up to form $\hat{A}^{RI}(\mathbf{r})$, therefore leading to no added singularities. 
However, in RI-OAT where data inconsistency is present, singularities whose extended line $l(\mathbf{r}, \xi)$ intersect the boundaries of $S_j$ may be missing in the reconstructed $\hat{A}^{RI}(\mathbf{r})$, and the added singularities may not cancel after summation. 

\subsection{Stable detection of singularities with insufficient illumination}
\label{subsec:stable_practical}
In practice, due to limited light penetration within a large object, the value of the optical fluence distribution $\Phi_j(\mathbf{r})$ within certain regions of the object may be negligibly small. 
In this case, there may be wavefront components of $\mu_a(\mathbf{r})$ that are, in effect, masked by these regions of $\Phi_j(\mathbf{r})$ that have negligible values. 
The optoacoustic signals originating from these regions cannot be reliably detected by the transducer. 
Under this \textit{insufficient illumination} condition, Assumption 3 can be regarded as being violated. 
In this subsection, by only utilizing Assumptions 1 and 2, we establish the microlocal correspondence for the RI-OAT problem with insufficient illumination. 

The wavefront set of $A_j(\mathbf{r})$ now satisfies: 
\begin{equation}
	WF[A_j(\mathbf{r})] \subset WF[\mu_a(\mathbf{r})], j = 1, 2, \dots, J. 
\end{equation}
The analysis in the previous subsection regarding the relationship between the visible and added singularities $\mathcal{V}_{S_j}(A_j(\mathbf{r})), \mathcal{A}_{S_j}(A_j(\mathbf{r}))$ and the estimated object $\hat{A}_j(\mathbf{r})$ for each limited-view subproblem remains valid (Eqn. (\ref{eq:A_j_thm})): 
\begin{equation}
	\mathcal{V}_{S_j}(A_j(\mathbf{r})) \subset WF[\hat{A}_j(\mathbf{r})] \subset \{ \mathcal{V}_{S_j}(A_j(\mathbf{r})) \cup \mathcal{A}_{S_j}(A_j(\mathbf{r}))\}. 
	\label{eq:A_j_thm_2}
\end{equation}
From Eqns. (\ref{eq:RI_summation}) and (\ref{eq:A_j_thm_2}), one obtains: 
\begin{multline}
	\cup_{j=1}^{J} \mathcal{V}_{S_j}(A_j(\mathbf{r})) \subset WF[\hat{A}^{RI}(\mathbf{r})] \\ 
	\subset (\cup_{j=1}^{J} \mathcal{V}_{S_j}(A_j(\mathbf{r}))) \cup (\cup_{j=1}^{J} \mathcal{A}_{S_j}(A_j(\mathbf{r}))). 
	\label{eq:union_thm_2}
\end{multline}
Equation (\ref{eq:union_thm_2}) indicates that $\hat{A}^{RI}(\mathbf{r})$ contains all visible singularities from the $J$ limited-angle subproblems, but will also possibly contain some added singularities attributable to any of the subproblems.
Note that unlike Eqn. (\ref{eq:union_thm}), Eqn. (\ref{eq:union_thm_2}) characterizes the singularities in $A_j(\mathbf{r})$ instead of $\mu_a(\mathbf{r})$. 
This is because in the insufficient illumination case, there are wavefront components in $WF[\mu_a(\mathbf{r})]$ that are masked and no longer present in $WF[A_j(\mathbf{r})]$.

Equation. (\ref{eq:union_thm_2}) reveals a microlocal correspondence for RI-OAT with insufficient illumination: 

\vspace{2ex}
\begin{assertion}
	\emph{
		In RI-OAT, under Assumptions 1 and 2, a wavefront set component $(\mathbf{r}, \xi)$ of $\mu_a(\mathbf{r})$ is present in the estimated $\hat{A}^{RI}(\mathbf{r})$, if and only if the followings holds: 
		\begin{enumerate}
			\item	The line $l(\mathbf{r}, \xi)$ extended in both directions by $\xi$ intersects with the interior of one of the measurement surface elements. Denote this surface element as $S_j$.  
			\item Denote the fluence distribution corresponding to $S_j$ as $\Phi_j(\mathbf{r})$. $\Phi_j(\mathbf{r})$ needs to be non-zero locally at $\mathbf{r}$, so that $(\mathbf{r}, \xi)$ will not be masked by the zero region of $\Phi_j(\mathbf{r})$ near $\mathbf{r}$. 
		\end{enumerate}
		Moreover, $\hat{A}^{RI}(\mathbf{r})$ may include additional artifacts, whose locations and directions are given by Eqn. (\ref{eq:added}) with $A(\mathbf{r}) = A_j(\mathbf{r})$.  
	}
\end{assertion}
\vspace{2ex}

\section{Computer-simulation studies}
\label{sec:method}
Computer-simulation studies were conducted to quantitatively investigate RI-OAT and to corroborate the theoretical conclusions made in Section \ref{sec:theoryCC}. 

\subsection{Illumination schemes}
\label{subsec:illum}
\begin{figure}[htbp]
	\centering
	\includegraphics[width=0.8\linewidth]{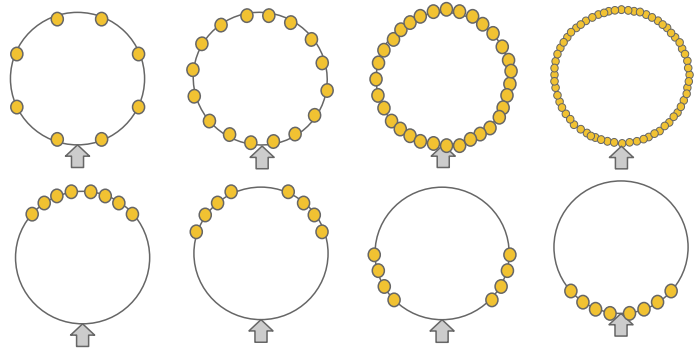}
	\caption{Illumination schemes used in the simulation studies. The yellow circles represent the positions of the illumination bars, which are positioned 8 cm from the center; the gray arrow represents the transducer location, which is also 8 cm from the center. Top row, from left to right, shows schemes 1 to 4, with 8, 16, 32, and 64 evenly spaced illumination bars, respectively. Bottom row, from left to right, shows schemes 5 to 8.} 
	\label{fig:illumSchemes}
\end{figure}

The eight implementations of 2D RI-OAT shown in Fig. \ref{fig:illumSchemes} were considered. 
In each case, the arrow denotes the position of the ultrasonic transducer relative to the light delivery system at a given tomographic view angle. 
\begin{description}
	\item [\textbf{Scheme 1 to Scheme 4:}]  The illumination bars were evenly spaced on a circle enclosing the object and delivered light towards the target to be imaged. 
		The number of illumination bars were 8, 16, 32, and 64 in schemes 1 - 4, respectively. 
		These schemes will be utilized to show how different degrees of data inconsistency affect the reconstructed images. 
	\item [\textbf{Scheme 5 to Scheme 8:}]  The number of illumination bars was fixed at 8, while the positions of the illumination bars varied in schemes 5 - 8. 
		In Scheme 5, the illumination bars were positioned on the opposite side of the transducer; 
		in Scheme 6, the illumination bars were positioned opposite to the transducer, but shifted to the flanks; 
		in Scheme 7, the illumination bars were positioned on the same side with the transducer, and shifted to the flanks; 
		in Scheme 8, the illumination bars were positioned adjacent to the transducer. 
		These schemes will be utilized to verify the microlocal correspondences in Section \ref{subsec:stable} and Section \ref{subsec:stable_practical} under different illumination set-ups. 
	\item [\textbf{Scheme 0: }] In addition, a situation in which the illumination was stationary and relatively uniform was considered. 
		In this scheme, 512 illumination bars were evenly spaced on a circle enclosing the object, and the illumination does not rotate during the imaging process. 
		This special scheme, referred to as \textbf{Scheme 0}, was employed to represent the conventional OAT. 
\end{description}

\subsection{Numerical phantoms}
\label{subsec:phantom}
\begin{figure}[htbp]
	\centering
	\includegraphics[width = 0.95\linewidth]{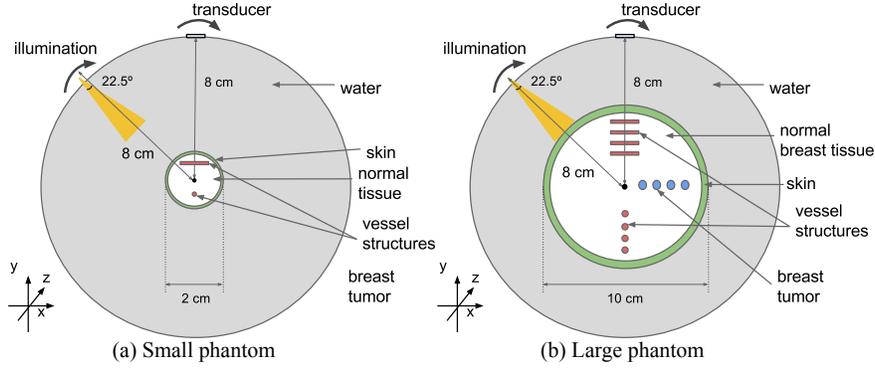}
	\caption{Slices through the central planes of the numerical phantoms.
	An illustration of the Monte Carlo (MC) simulation set-up is also displayed. 
	(a) The small phantom to mimic sufficient illumination (small animal imaging). 
	(b) The large phantom to mimic insufficient illumination (breast imaging). Only one illumination bar is shown for illustration purpose, while the actual MC simulation employed multiple illumination bars.}
	\label{fig:NumericalPhantom}
\end{figure}

Two 3D numerical phantoms were employed. 
Slices through the central planes of the phantoms are shown in Fig. \ref{fig:NumericalPhantom}. 
\begin{description}
	\item [\textbf{Small Phantom:}]  The phantom on the left in Fig. \ref{fig:NumericalPhantom} corresponds to a relatively small object that is representative of small animal imaging. 
		The torso is represented by a 2 cm diameter cylinder with 1 mm thick skin; vessels are represented by two small cylinders oriented in the X-Y and Y-Z directions. 
		This phantom simulates the scenario in Section \ref{subsec:stable}, where the delivered light can sufficiently penetrate through the object such that Assumptions 1, 2, and 3 are all satisfied. 
	\item [\textbf{Large Phantom:}]  The phantom on the right in Fig. \ref{fig:NumericalPhantom} corresponds to a larger object that is representative of clinical breast imaging. 
		The breast is represented by a 10 cm diameter cylinder with an 1 mm thick skin; vessels are represented by eight small cylinders oriented in the X-Y and Y-Z directions; four spheres represent tumors. 
		This phantom simulates the insufficient illumination scenario in Section \ref{subsec:stable_practical} where, due to limited light penetration, the light fluence at certain locations is negligible.
\end{description}

\subsection{Generation of simulated tomographic measurements}
\subsubsection{Monte Carlo simulation of optical process}
\label{subsec:MC_simulation}
To simulate the light propagation process, a GPU-accelerated 3D Monte Carlo method was employed \cite{MCXToolbox, JacquesMCPaper}. 
In the MC simulation, the total volume is $162.5 \times 162.5 \times 7.75 \rm{ mm}^3$, with a voxel-size of 0.25 mm. 
The incident light from each illumination bar was modeled as a uniform cone beam with an angle of 22.5 degrees located at the center of the Z direction. 
Light propagation through both numerical phantoms in Section \ref{subsec:phantom} were simulated, with representative optical property values assigned to different tissues within the phantom (see Appendix \ref{append:OpticalProperty}). 
The Monte Carlo simulation was repeated for each of 512 tomographic view angles to obtain the corresponding view-dependent 3D optical absorption distribution. 
Finally a thin 3D slice along the X-Y plane with a Z-direction height of 0.25 mm was extracted to represent the object function $A_j(\mathbf{r})$. 
The 3D thin slice was employed instead of the entire 3D volumetric optical absorption distribution in order to reduce computation burden. 
In addition, the thin 3D slice well resembles a 2D scenario which can greatly simplify the analysis in Section \ref{sec:simulation}. 
This process was repeated for all eight RI-OAT illumination schemes in Section \ref{subsec:illum} alone with Scheme 0. 

\subsubsection{Acoustic pressure data generation}
\label{subsec:forward_data}
Acoustic wave propagation within the 2D plane resembled by the 3D slice $A_j(\mathbf{r})$ was simulated. 
A $1.10 \times 1.10 \mathrm{ mm}^2$ planar transducer was employed to record the pressure signal. 
One complete scan corresponded to rotating both the illumination and transducer for $J = 512$ steps around the object, with the transducer facing the center of rotation. 
At every step, the initial pressure distribution was set to $\Gamma A_j(\mathbf{r})$.
The pressure data recorded at the transducer location were generated by the spherical-voxel-based-method described in Eqn. (\ref{eq:discrete_basic}) with the system matrix $\mathbf{H}$ computed by use of Eqn. (\ref{eq:system_matrix}) (details can be found in \cite{KenjiPaper}). 
The sampling frequency was 10 MHz, with $K=1024$ time samples and $L = 1024$ frequency samples. The radius of the spherical voxel was $\epsilon = 0.125$ mm. The speed of sound was $c = 1.52$ mm$/ \mathrm{\mu}$s. 
The EIR of the transducer is set to be a Dirac-delta function, and the SIR was modeled using Eqn. (\ref{eq:SIR}). 

\subsection{Image reconstruction methods}
\label{subsec:image_recon_method}

A FBP and an iterative reconstruction method for 3D OAT were employed to reconstruct images from the simulated RI-OAT pressure data. 
We applied 3D reconstruction methods because the thin 3D object produces acoustic wave-fields that obey a 3D wave equation. 
Both methods were predicated upon the assumption that the object and background medium were acoustically homogeneous and lossless.  This assumption is routinely employed with good success in current implementations of OAT \cite{Yang_Mouse, KunMouse, wang2011imaging, KenjiPaper}. 

The FBP method corresponded to a discretized form of the following 3D reconstruction formula \cite{WangBook}:
\begin{equation}
	A(\mathbf{r}) = \frac{C_p}{2\pi \beta c^2} \int_{S} \frac{\hat{\mathbf{n}_0^s} \dot (\mathbf{r'} - \mathbf{r}) d S_0}{|\mathbf{r} - \mathbf{r}'|^3} \bigl[ p(\mathbf{r'}, t) - 2 t\frac{\partial p(\mathbf{r}', t)}{\partial t} \bigr] |_{t = \frac{|\mathbf{r} - \mathbf{r}'|}{c}}. 			
	\label{eq:fbp}
\end{equation}
Here, $S$ denotes the measurement surface, $d S_0$ denotes the detection element, and $\hat{\mathbf{n}_0^s}$ denotes the unit normal vector of $d S_0$ pointing inward. 
Although it can be implemented efficiently\cite{KunGPU}, like most analytic reconstruction formula, (\ref{eq:fbp}) is based on the canonical C-C imaging model that assumes idealized point-like transducers and complete measurement data.

Because they are directly formulated on a D-D imaging model instead of the canonical C-C model, iterative, also known as optimization-based, image reconstruction methods provide the opportunity to compensate for non-ideal physical factors and better mitigate data incompleteness.
In this study, based on the D-D model in Eqn. (\ref{eq:discrete_basic}), a penalized least-squares (PLS) estimate $\boldsymbol{\theta}_{opt}$ was defined as
\begin{equation}
	\boldsymbol{\theta}_{opt} = \argmin_{\boldsymbol{\theta}} || \mathbf{\tilde{u}} - \mathbf{H}\boldsymbol{\theta} ||_2^2	+ \lambda R(\boldsymbol{\theta}) .
	\label{eq:optm}
\end{equation}
Here, $R(\boldsymbol{\theta})$ is a regularization penalty function whose effect is controlled by the regularization parameter $\lambda  \in \mathbb{R}$. The following quadratic penalty was employed \cite{KunGPU}:
\begin{equation}
	R(\boldsymbol{\theta}) = \sum_{n=0}^{N-1} ([\boldsymbol{\theta}]_n - [\boldsymbol{\theta}]_{n_x})^2 + ([\boldsymbol{\theta}]_n - [\boldsymbol{\theta}]_{n_y})^2 + ([\boldsymbol{\theta}]_n - [\boldsymbol{\theta}]_{n_z})^2 ,
	\label{eq:reg}
\end{equation}
where $n_x$, $n_y$, and $n_z$ denote the index values of the neighboring voxels of $[\boldsymbol{\theta}]_n$ along the $X$, $Y$, and $Z$ directions.
This iterative method will hereafter be referred to as the PLS method. 
The system matrix $\mathbf{H}$ was computed by use of Eqn. (\ref{eq:system_matrix}), which models the EIR and SIR of the transducer element. 
Although the system matrix employed in the reconstruction was the same as the one employed in the forward data generation in Section \ref{subsec:forward_data}, exact inverse crime was avoided because of the aforementioned data inconsistency caused by view-dependent $A_j(\mathbf{r})$ in RI-OAT.
A conjugate gradient (CG) algorithm \cite{KunGPU} was employed to (approximately) solve Eqn. (\ref{eq:optm}). 
The algorithm was terminated after 100 iterations.
The regularization parameter $\lambda$ was empirically chosen to avoid over-smoothing the reconstructed image.

For both FBP and PLS methods, the 3D region to be reconstructed was of size $30 \times 30 \times 0.25 \text{ mm}^3$ for the small phantom, and $110 \times 110 \times 0.25 \text{ mm}^3$ for the large phantom, both with an voxel-size of 0.25 mm. 
The Gruneisen coefficient was $\Gamma = \beta c^2 / C_p = 2000$ in all studies.

\section{Numerical results}
\label{sec:simulation}

\subsection{Comparison of different reconstruction methods}
\label{subsec:recon_method_compare}
To investigate the effects of different physical factors on RI-OAT image reconstruction, three different sets of simulated pressure data were generated: 
\begin{enumerate}
	\item Pressure data recorded by an ideal point-like-transducer: This dataset was employed to investigate the reconstruction methods' robustness to data inconsistency produced by the use of rotating illumination. 
	\item Pressure data produced in the same way as above, but with white Gaussian noise added. 
		The standard deviation of the noise is 1\% of the peak value of the noiseless measurement data. 
	\item Pressure data recorded by a finite-size-transducer with measurement noise: White Gaussian noise with a standard deviation of $1\%$ of the peak value of the measurement data was added, and the transducer's SIR was incorporated in the forward model. This dataset was employed to investigate the reconstruction method's performance in a more realistic situation. 
\end{enumerate}
Note that because the FBP method does not compensate for the transducer's directivity, point-like-transducer is assumed in the first two sets of forward data to minimize the effect of additional model error, while a finite-size-transducer is assumed for the third set of forward data. 
For simplicity, we show the results for illumination Scheme 5 only.

\begin{figure}
	\centering
	\includegraphics[width = \linewidth]{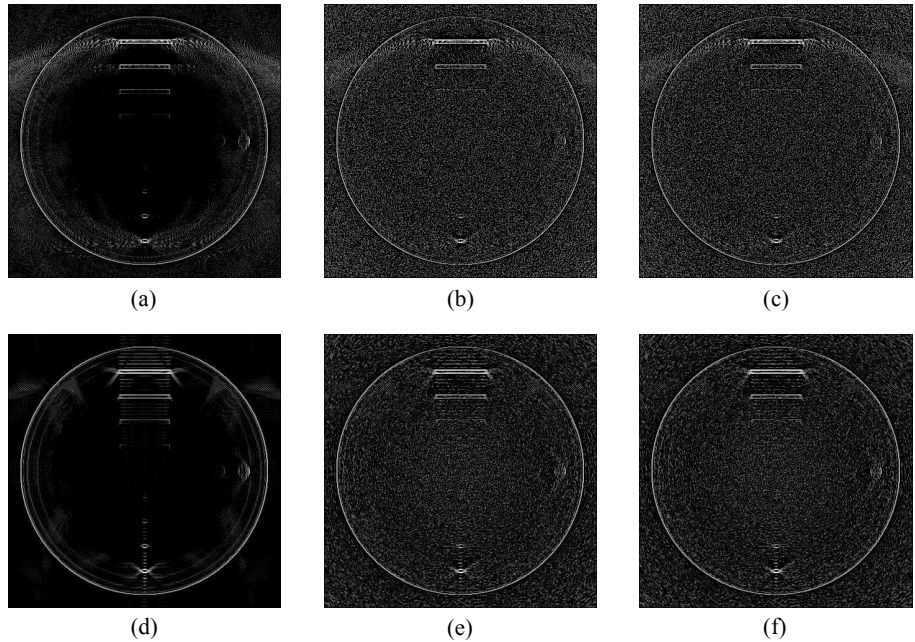}
	\caption{Comparison of FBP and PLS reconstruction methods with three sets of forward data for Scheme 5. All images are displayed in logarithmic scale, with the same gray scale window showing black as the minimum value and white as the maximum. (a) FBP with point-transducer, noiseless. (b) FBP with point-transducer, noisy. (c) FBP with finite-size-transducer, noisy. (d) PLS with point-transducer, noiseless. (e) PLS with point-transducer, noisy. (f) PLS with finite-size-transducer, noisy. }
	\label{fig:recon_method_compare}
\end{figure}

Both the FBP and the PLS reconstruction methods were applied to all three sets of simulated measurement data. 
The reconstructed images are shown in Fig. \ref{fig:recon_method_compare}, where the top row shows the FBP results for the three datasets, and the bottom row contains the corresponding PLS results. 
Figure \ref{fig:recon_method_compare}(a) shows significant structured streak-type artifacts, likely caused by model inconsistencies from rotating illumination, while these artifacts are largely suppressed in (d). 
Figure \ref{fig:recon_method_compare}(b) shows random background noise in addition to the inconsistency-artifacts, while a cleaner background in (e) suggests successful noise suppression by the PLS method. 
A comparison between (c) and (f) also demonstrates PLS method's ability to compensate for transducer directivity while mitigating various types of artifacts.
These results indicate that, compared to the FBP method, the PLS method can better compensate for artifacts caused by data inconsistency from rotating illumination and measurement noise, and transducer directivity. 
Therefore, in the following two studies the PLS method was employed.

\subsection{Relationship between data inconsistency and image degradation}
\label{subsec:recon_1_4}
\begin{figure}[htbp]
	\centering
	\includegraphics[width = \linewidth]{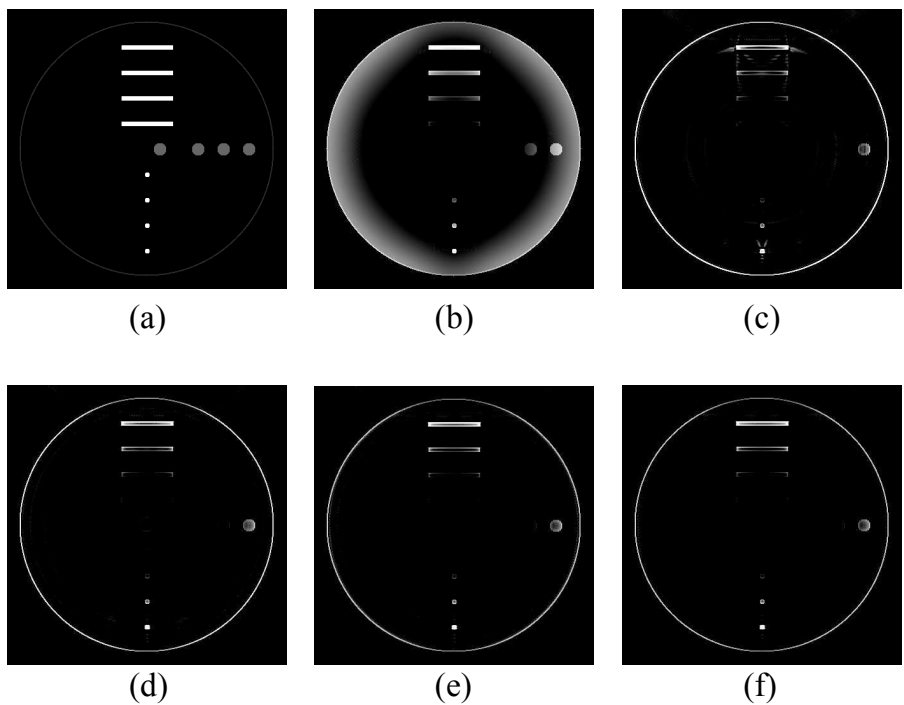}
	\caption{Image reconstructed by use of the PLS method for Scheme 1 through Scheme 4, displayed in logarithmic scale. (a) phantom absorption coefficient $\mu_a$ map, (b) Scheme 0 (stationary illumination), (c) Scheme 1, (d) Scheme 2, (e) Scheme 3, (f) Scheme 4. }
	\label{fig:recon_1_4}
\end{figure}

Figure \ref{fig:recon_1_4} shows images reconstructed by use of the PLS method corresponding to the large phantom with illumination schemes 1 to 4. 
Table \ref{tab:error_1} provides the mean square error (MSE) between the reconstructed image and the phantom absorption map. 
Since $A_j(\mathbf{r})$ is different at each view in RI-OAT, we do not have a ground truth $A(\mathbf{r})$. 
Thus, we displayed the optical absorption coefficient map $\mu_a(\mathbf{r})$ as a reference in Fig. \ref{fig:recon_1_4}(a), and used the average $A(\mathbf{r})$ of all $512$ views from the Monte Carlo simulation (similar to Eqn. (\ref{eq:RI_summation})) as the ground truth $A(\mathbf{r})$ when computing the MSE. 
We also displayed the reconstructed image from Scheme 0 in Figure \ref{fig:recon_1_4}(b) corresponding to stationary illumination. 
The results show a direct relationship between the degree of inconsistency and the degree of the reconstructed image's degradation. 

\begin{table}
	\renewcommand{\arraystretch}{1.3}
	\centering
	\begin{threeparttable}[htbp]
		\caption{Mean Square Error Between Reconstructed Images and Phantom Absorption Maps for Large phantom with schemes 1 through 4}
		\begin{tabular}{l | c c c c}
				\hline
				~ 											&Scheme 1 	&Scheme 2		&Scheme 3		&Scheme 4		\TStrut \BStrut \\	\hline
				MSE\tnote{1}						&6.0377			&2.8530			&2.3990			&2.3867			\TStrut \BStrut \\ \hline 
		\end{tabular}

		\begin{tablenotes}
		\item[1] The mean square error is in scales of $\times 10^{-4}$.
		\end{tablenotes}
		\label{tab:error_1}
	\end{threeparttable}
\end{table}

When fewer illumination bars are utilized (as in Scheme 1), $A_j(\mathbf{r})$ changes significantly when $j$ varies, which leads to a large degree of data inconsistency between the RI-OAT imaging model and the stationary OAT imaging model assumed by the reconstruction algorithm. 
When the number of illumination bars is increased (in schemes 2 to 4), the difference in $A_j(\mathbf{r})$ for different $j$ becomes smaller, hence lowering the degree of data inconsistency. 
With a smaller degree of inconsistency, the reconstruction algorithm can better recover structures within the object. 
In Figure \ref{fig:recon_1_4}, one can observe visible blurring and streak-type artifacts in Scheme 1, but these artifacts become less severe when the illumination bar number increases. 
In Table \ref{tab:error_1}, a decrease in the MSE with an increased number of illumination bars is also observed. 
These results confirm that the degree of data inconsistency in RI-OAT is directly related to the degree of degradation in the reconstructed images.

\subsection{Verification of singularity detection with sufficient illumination}
\label{subsec:verify_ideal}
\begin{figure}[htbp]
	\centering
	\includegraphics[width= 0.8 \linewidth]{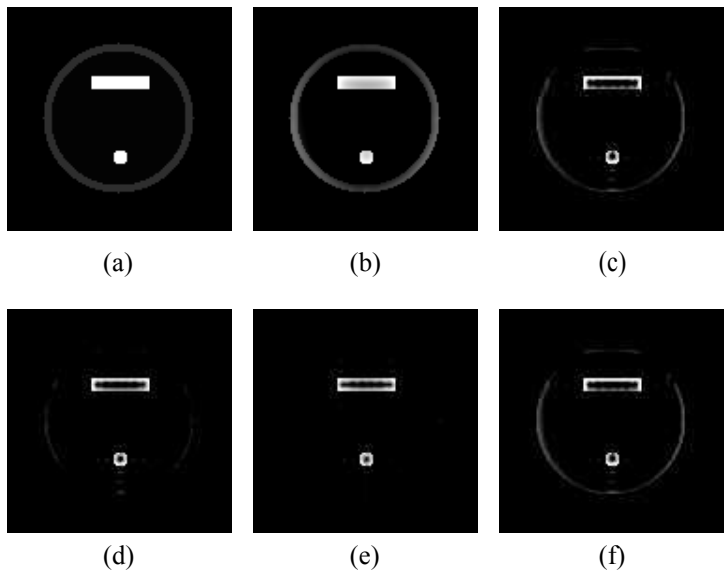}
	\caption{Reconstructed images of the small phantom for schemes 5 through 8, displayed in logarithmic scale. (a) phantom absorption coefficient $\mu_a$ map, (b) Scheme 0 (stationary illumination), (c) Scheme 5, (d) Scheme 6, (e) Scheme 7, (f) Scheme 8. }
	\label{fig:recon_5_8_small}
\end{figure}
Here, results that corroborate the singularity detection condition (Assertion 1) in RI-OAT with sufficient illumination are presented. 
Illumination schemes 5 through 8, which represents very different illumination conditions, were applied with the small phantom described in Section \ref{subsec:phantom}. 
The results for Scheme 0 are presented as a reference.
The reconstructed images are shown in Fig. \ref{fig:recon_5_8_small}. 
Because the wavefront set is applicable only to continuous functions, we use the edges of a discrete image to approximate its `singularities'. 
Here, the edges were extracted by applying a Roberts Cross filter \cite{roberts} to the discrete image to estimate the gradient magnitude of the image, and thresholding the magnitude of the gradient map with a threshold of 3 times of the image's mean intensity value. This value represents a practical threshold below which the signal is assumed to be too small to detect. 
Note that for Scheme 0, we used a different threshold of 1 times the mean of the image's intensity value in order to get a representative edge map (see Section \ref{sec:discussion} for more discussion on the choice of threshold). 

\begin{figure}[htbp]
	\centering
	\includegraphics[width = \linewidth]{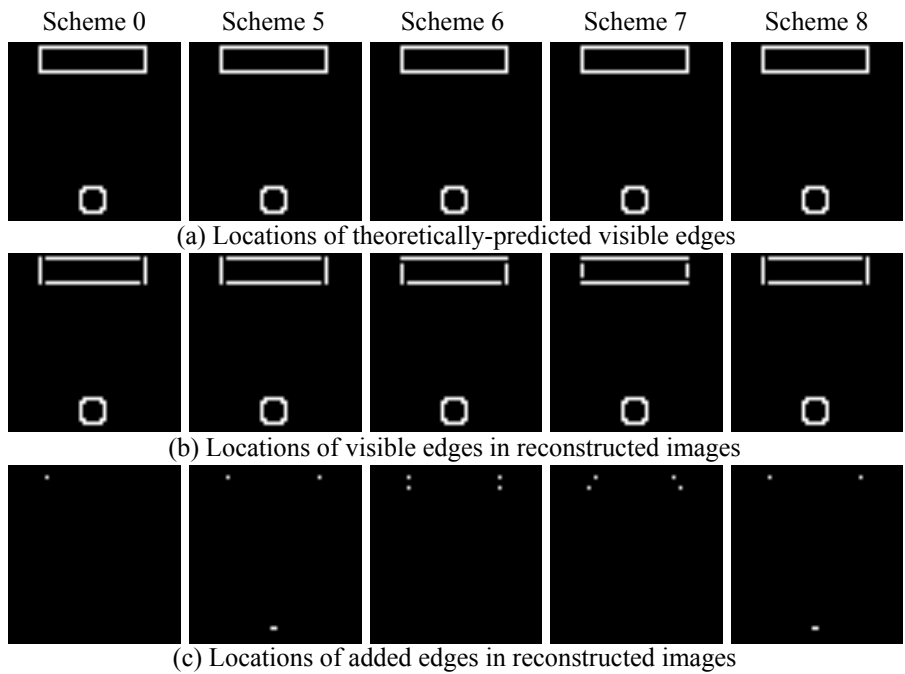}
	\caption{Corroboration of Assertion 1 for sufficient illumination. The columns from left to right in each sub-figure display the results for schemes 0, 5, 6, 7, and 8, respectively. (a) the stably detectable (visible) edges predicted by the corollary. (b) the reconstructed visible edges from OAT simulation. (c) the reconstructed added edges from OAT simulation. }
	\label{fig:theory_compare_small}
\end{figure}

To corroborate Assertion 1, the following steps were performed, with the results shown in Fig. \ref{fig:theory_compare_small}: 
\begin{enumerate}
	\item First, the edges of the absorption map $\mu_a(\mathbf{r})$ were extracted using the aforementioned Roberts edge detection method. 
	An area of $12.5 \times 12.5 \text{mm}^2$,  which contains the tube and circle structures within the phantom, was cropped from the reconstruction region. 
	\item Assertion 1 was employed to predict the visible singularities in $\mu_a(\mathbf{r})$, which are shown in Fig. \ref{fig:theory_compare_small}(a). 
	\item The Roberts edge detection method was applied to the reconstructed images with the same noise threshold to retrieve the edge maps. 
		The reconstructed images contain both the visible singularities and added singularities. 
		We separated them using a mask generated from the $\mu_a(\mathbf{r})$ edge map in Step 1. 
		The reconstructed visible singularities were displayed in Fig. \ref{fig:theory_compare_small}(b), and the reconstructed added singularities were displayed in Fig. \ref{fig:theory_compare_small}(c). 
	\item Finally, the mean square error (MSE) and structure similarity (SSIM) index \cite{ssim} were computed between the theoretical edge maps obtained in Step 2 and the reconstructed visible edge maps obtained in Step 3, which are shown in Table \ref{tab:error_small} for reference. 
\end{enumerate}
As stated in section \ref{subsec:stable}, for RI-OAT with sufficient illumination and a closed detection surface, all the singularities in $\mu_a(\mathbf{r})$ can be stably recovered, as indicated by the results in Fig. \ref{fig:theory_compare_small}(a). 
This conclusion is further corroborated by the observation that almost all the edges are accurately reconstructed in the simulation results shown in Fig. \ref{fig:theory_compare_small}, and almost no added singularities are present in Fig. \ref{fig:theory_compare_small}(c). 
Along with the quantitative measures in Table \ref{tab:error_small}, the close agreement between the theoretical prediction and simulation results corroborates Assertion 1 regarding the stable detection of singularities in RI-OAT with sufficient illumination. 

\begin{table}[htbp]
	\renewcommand{\arraystretch}{1.3}
	\centering
	\caption{Similarities of visible edges between theoretical predictions and reconstructed images}
	\begin{tabular}{l | c c c c}
		\hline
		~					  &Scheme 5 	&Scheme 6		&Scheme 7		&Scheme 8		\TStrut \BStrut	\\	\hline
		MSE					&0.0015379	&0.0015379	&0.0015379	&0.0015379		\TStrut	\\
		SSIM				&0.99963		&0.99954		&0.99960		&0.99960		\BStrut	\\	\hline
	\end{tabular}
	\label{tab:error_small}
\end{table}

\subsection{Verification of singularity detection with insufficient illumination}
\label{subsec:verify_practical}
\begin{figure}[htbp]
	\centering
	\includegraphics[width=\linewidth]{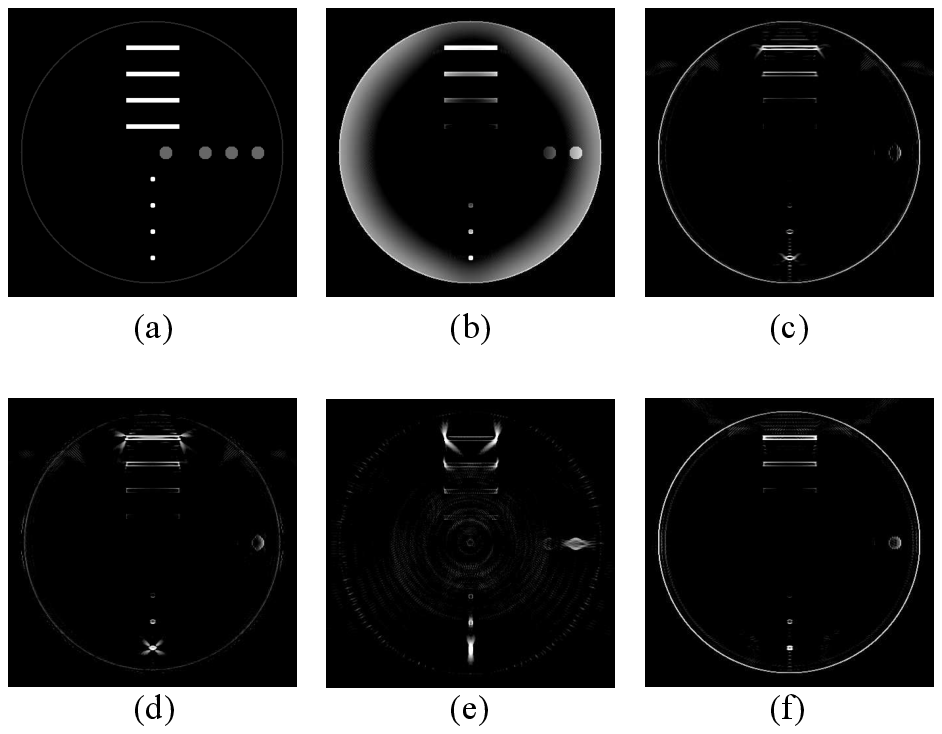}
	\caption{Reconstructed images for large phantom with Scheme 5 through Scheme 8, displayed in logarithmic scale. (a) phantom absorption coefficient $\mu_a$ map; (b) Scheme 0 (stationary illumination); (c) Scheme 5; (d) Scheme 6; (e) Scheme 7; (f) Scheme 8. }
	\label{fig:recon_5_8}
\end{figure}
\begin{figure}[htbp]
	\centering
	\includegraphics[width = 0.6\linewidth]{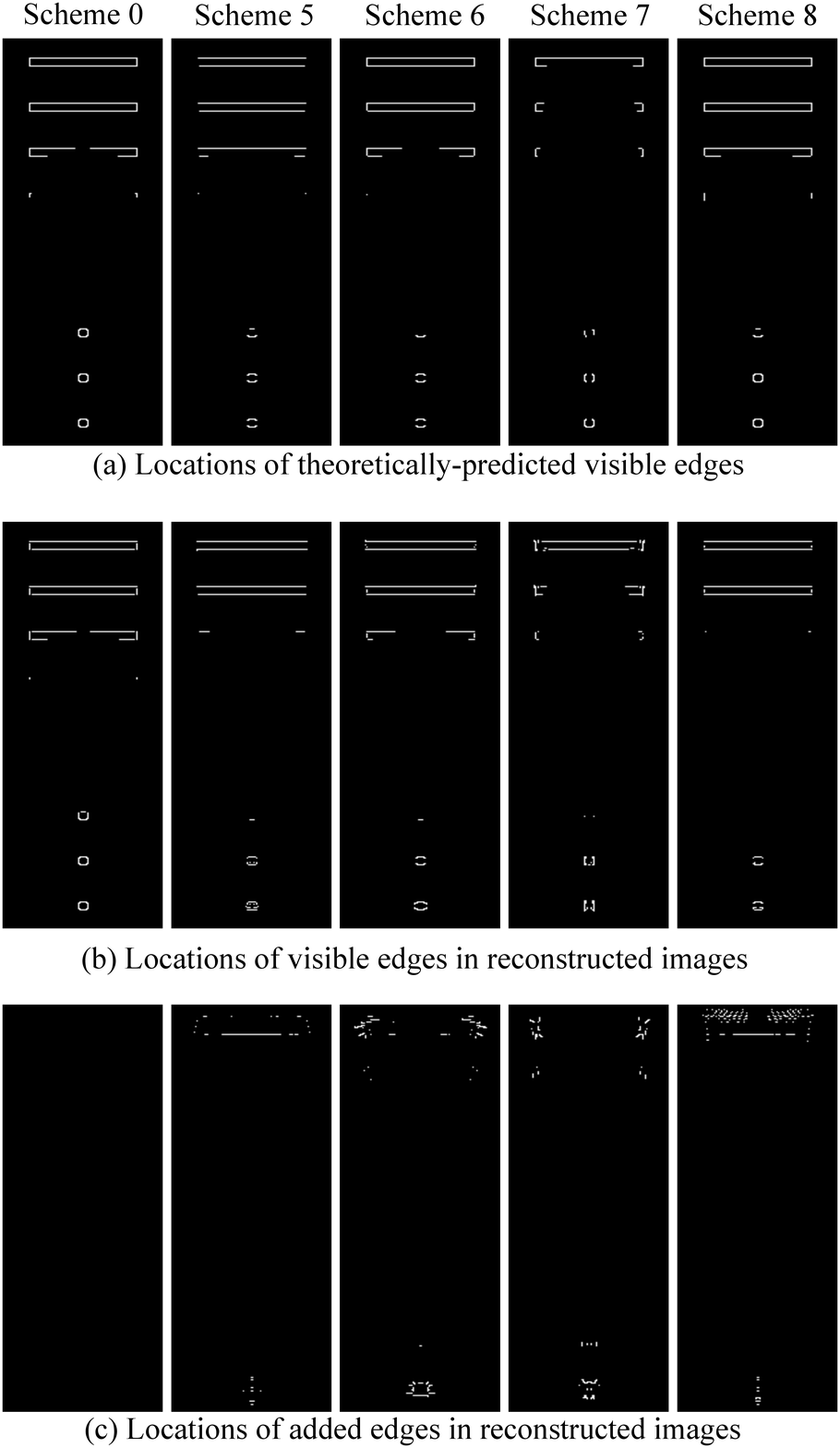}
\caption{Corroboration of Assertion 2 for insufficient illumination.  The columns from left to right in each sub-figure display results for schemes 0, 5, 6, 7 and 8, respectively. (a) the stably detectable (visible) edges predicted by Assertion 2. (b) the reconstructed visible edges from OAT simulation. (c) the reconstructed added edges from OAT simulation. }
	\label{fig:theory_compare}
\end{figure}

In this subsection, images of the large phantom corresponding to illumination schemes 5-8 were reconstructed from the simulated pressure data. 
Scheme 0 was also considered to represent stationary illumination as a reference. 
The reconstructed images are shown in Fig. \ref{fig:recon_5_8}. 
The same 4-step procedure described in the previous section was employed to verify the singularity detection condition with insufficient illumination (Assertion 2). 
The threshold of Roberts edge detection is chosen to be 4 times the mean intensity of the image. 
The results are shown in Fig. \ref{fig:theory_compare} and in Table \ref{tab:error_2}. 

In the continuous-to-continuous context where finite sampling effects are neglibigle, with stationary illumination and the measurement surface enclosing the object, all singularities in the optical absorption map will be stably recovered in the reconstructed image, which is corroborated by Scheme 0 in Fig. \ref{fig:theory_compare}(a) and \ref{fig:theory_compare}(b). 
In addition, no added singularity will be present in the reconstructed image, which is demonstrated by Scheme 0 in Fig. \ref{fig:theory_compare}(c). 
And for RI-OAT, as stated in Assertion 2, even with a closed measurement surface, singularities in the object may be missing in the reconstructed image if they are located within the zero regions of the fluence map when light penetration is limited. 
This is corroborated by the agreement between the theoretical prediction of visible singularities (Fig. \ref{fig:theory_compare}(a)) and the reconstructed visible singularities (Fig. \ref{fig:theory_compare}(b)). 
Assertion 2 also predicts that some additional singularities may be present in the reconstructed image, which is confirmed by Fig. \ref{fig:theory_compare}(c). 
Note that the theoretical analysis in section \ref{sec:theoryCC} is based on a mathematical and continuous context, which is inherently different from the discrete framework on which the simulation studies are based, therefore the agreement between Fig. \ref{fig:theory_compare}(a) and Fig. \ref{fig:theory_compare}(b) may not be perfect. 
Despite this discrepancy, the close resemblance between the theoretical prediction and simulation results still implies that Assertion 2 provides useful guidance for determining missing and added singularities in reconstructed images in RI-OAT with insufficient illumination. 

\begin{table}[htbp]
	\renewcommand{\arraystretch}{1.3}
	\centering
	\caption{Similarities of visible edges between theoretical predictions and reconstructed images}
	\begin{tabular}{l | c c c c}
		\hline
		~					  &Scheme 5 	&Scheme 6		&Scheme 7		&Scheme 8		\TStrut \BStrut	\\	\hline
		MSE					&0.00364		&0.00392		&0.00394		&0.00267		\TStrut	\\
		SSIM				&0.99986		&0.99985		&0.99989		&0.99990		\BStrut	\\	\hline
	\end{tabular}
	\label{tab:error_2}
\end{table}

\section{Discussion and conclusion}
\label{sec:discussion}
This paper investigated the impact of non-stationary illumination on OAT image reconstruction. 
Based on the canonical OAT imaging model, the imaging model of RI-OAT was described in the continuous form. 
Theoretical insights into the RI-OAT reconstruction problem were provided by applying mathematical results from microlocal analysis. 
Theoretical assertions were proposed to identify the visible and added singularities within the object for RI-OAT under both sufficient and insufficient illumination conditions. 
The designed numerical simulation studies showed that in RI-OAT: 
1) The degree of data inconsistency is directly related to the degree of reconstructed image degradation. 
2) Iterative image reconstruction methods can better mitigate reconstruction artifacts caused by both data inconsistency and noise, compared to analytical methods like filtered back-projection. 
3) The singularities stably reconstructed in the simulation closely match those predicted by our theoretical conclusions under various illumination conditions. 

The studies in Sections \ref{subsec:verify_ideal} and \ref{subsec:verify_practical} compared the theoretically-predicted visible singularities, $\mathcal{V}(\mu_a(\mathbf{r}))$, with the visible singularities contained in the images reconstructed by a reconstruction method that assumed a stationary optical illumination. 
Although Eqn. (\ref{eq:added}) defines the set of added singularities, $\mathcal{A}(\mu_a(\mathbf{r}))$, for each tomographic view angle, when the estimated $\hat{A}_j(\mathbf{r})$ are summed up over all tomographic views, these added singularities may cancel (see Eqn. (\ref{eq:RI_summation})). 
The degree of data inconsistency, assuming an idealized measurement system, increased from zero in the stationary illumination case, to a relatively small degree in RI-OAT with sufficient illumination case, and further increases in RI-OAT with insufficient illumination. 
The added singularities will completely cancel out with stationary illumination under a continuous context (see Scheme 0 in Fig. \ref{fig:theory_compare_small}). 
The presence of more streak-type artifacts in Fig. \ref{fig:recon_5_8} than in Fig. \ref{fig:recon_5_8_small} suggests less cancellation in the insufficient illumination case compared to the sufficient illumination case.     
Therefore, it is reasonable to assume that a larger degree of data inconsistency results in less cancellation of added singularities, and thus more artifacts in the reconstructed image. 

The threshold value used in the Roberts edge detection step in Section \ref{sec:simulation} determined the value of $A(\mathbf{r})$ under which the absorbed optical energy was assumed to be negligible. 
Developing a rigorous threshold-picking-method is beyond the scope of this work. 
We chose a threshold based on the mean intensity value of the image, and visually verified that the extracted edge map is a close representation of the edges in the image.

\section*{Funding Information}
This work was supported in part by NIH awards CA1744601 and EB01696301.

The authors thank Dr. Linh Nguyen and Dr. Todd Quinto for useful conservations regarding the application of microlocal analysis to optoacoustic tomography.

\section*{Appendices}
\subsection{Formal definition of a wave front set}
\label{append:wave_front_set}
The formal definition of a wave front set and additional background information on microlocal analysis is provided below \cite{Quinto_LimitAngle}. 

Denote the space of compactly supported smooth functions as $ \mathcal{D}(\mathbb{R}^N)$ and its dual space as $\mathcal{D}'(\mathbb{R}^N)$. Further denote the space of functions that are differentiable for all degrees as $\mathcal{E}(\mathbb{R}^N) = C^{\infty}(\mathbb{R}^N)$ and its dual space as $\mathcal{E}'(\mathbb{R}^N)$. Next we define \emph{smooth} and \emph {decay rapidly}.

\vspace{2ex}\noindent{\footnotesize {\bf Definition 1} 
	A function $A(\mathbf{r})$ is said to be \textbf{smooth} if it is in $\mathcal{E}(\mathbb{R}^N)$. 
}

\vspace{2ex}\noindent{\footnotesize {\bf Definition 2} 
	A smooth function $A(\mathbf{r})$ is said to \textbf{decay rapidly} on a conic open set $V$ if, for any integer $n$, there is a constant $C_n$ such that $|A(\mathbf{r})| \leq C_n (1+ |\mathbf{r}|)^{-n}, \forall \mathbf{r} \in V$. 
}

Define \emph{the singular support} of a function $A(\mathbf{r})$, $\sing \supp(A)$ as the complement of the largest open set on which $A(\mathbf{r})$ is in $C^\infty$. Define \emph{frequency set} as follows \cite{Quinto_LimitAngle}: 

\vspace{2ex}\noindent{\footnotesize {\bf Definition 3}
	Let $A(\mathbf{r}) \in \mathcal{E}'(\mathbb{R}^N)$. We define the \textbf{frequency set} $\Sigma(A)$ of $A(\mathbf{r})$ as the set of all directions $\xi \in \mathbb{R}^N \setminus 0$ in which $\mathcal{F}(A)$ does not decay rapidly in any conic neighborhood of $\xi$ ($\mathcal{F}$ denotes Fourier transform). 
}

For $A(\mathbf{r})$, $\sing \supp(A)$ gives the location of the singularities in $A(\mathbf{r})$, and the frequency set $\Sigma(A)$ describes all directions along which $A(\mathbf{r})$ is singular. By introducing a cutoff function $\varphi \in \mathcal{D}(\mathbb{R}^N)$, we can simultaneously describe the location and direction of a singularity: 

\vspace{2ex}\noindent{\footnotesize {\bf Definition 4}
	Let $A(\mathbf{r}) \in \mathcal{D}'(\mathbb{R}^N)$. The \textbf{localized frequency set} of $A$ at $\mathbf{r} \in \mathbb{R}^n$ is defined as 
	\begin{equation}
		\Sigma_r(A) = \bigcap {\Sigma(\varphi A): \varphi \in \mathcal{D}(\mathbb{R}^N), \varphi(\mathbf{r}) \neq 0}. 
	\end{equation}
}
Then the localized frequency set $\Sigma_r(A)$ is the set of directions along which $A(\mathbf{r})$ is singular at $\mathbf{r}$. This gives the formal definition of a wave front set: 

\vspace{2ex}\noindent{\footnotesize {\bf Definition 5}
	Let $A(\mathbf{r}) \in \mathcal{D}'(\mathbb{R}^N)$. The \textbf{wave front set} of $A(\mathbf{r})$ is given by: 
	\begin{equation}
		WF[A] = {(\mathbf{r}, \xi) \in \mathbb{R}^N \times {\mathbb{R}^N \setminus 0}, \text{where: } \mathbf{r} \in \sing \supp{A}, \xi \in \Sigma_r{A}. }
	\end{equation}
}

From Theorem 11 in \cite{Brouder_wavefrontset}, a function $A(\mathbf{r})$ is smooth if and only if its wave front set is empty.

\subsection{Light propagation model of OAT}
\label{append:light}
In OAT, an incident light source is employed to irradiate the object, and the absorbed optical energy produces an acoustic wave field via the photoacoustic effect. 
The light propagation process can be modeled by the radiative transfer equation (RTE) \cite{WangBook}: 
\begin{multline}
	\frac{\partial L(\mathbf{r}, \hat{s}, t)}{c \partial t} = -\hat{s} \nabla L(\mathbf{r}, \hat{s}, t) - (\mu_a(\mathbf{r}) + \mu_s(\mathbf{r})) L(\mathbf{r}, \hat{s}, t)		\\ 
	+ \mu_s \int_{4 \pi} L(\mathbf{r}, \hat{s}', t) P(\hat{s}', \hat{s}) d \Omega' + S(\mathbf{r}, \hat{s}, t), 
	\label{eq:RTE}
\end{multline}
where $L(\mathbf{r}, \hat{s}, t)$ is the radiance, $\hat{s}$ denotes unit direction vector, $c$ is the speed of sound, $\mu_a(\mathbf{r})$ and $\mu_s(\mathbf{r})$ denotes the optical absorption and scattering coefficients, and $\nabla$ is the gradient operator.
The product $P(\hat{s}', \hat{s}) d \Omega$ denotes the probability of light with propagation direction $\hat{s}'$ being scattered into a small solid angle range $d \Omega$ around direction $\hat{s}$, and $S(\mathbf{r}, \hat{s}, t)$ denotes the incident light source. 
The optical absorption process is then described by the following equation: 
\begin{equation}
	A(\mathbf{r}) = \mu_a(\mathbf{r}) \int dt \int_{4 \pi} L(\mathbf{r}, \hat{s}, t) d \Omega, 
	\label{eq:AofR}
\end{equation}
where the compactly supported and bounded function $A(\mathbf{r})$ represents the absorbed optical energy density. 
The conversion from $A(\mathbf{r})$ to the initial pressure field is further described by Eqn. (\ref{eq:wave_condition}). 
In practice, the Monte Carlo (MC) method is considered numerically equivalent to the RTE model. 
Therefore, a 3D MC method was employed in our study to simulate the light propagation process. 

\subsection{RI-OAT imaging model and interpretation: Discrete case}
\label{append:discrete}
In this section, the imaging model for RI-OAT is presented in its discrete form.  
This model is employed to provide a physical interpretation of an image reconstructed image by use of a conventional linear OAT reconstruction operator that assumes a stationary light fluence distribution. 

\subsubsection{D-D imaging model for RI-OAT}
Recall the conventional D-D OAT imaging model in Eqn. (\ref{eq:discrete_basic}): 
\begin{equation}
	\tilde{\mathbf{u}} = \mathbf{H} \boldsymbol{\theta}. 		
	\label{eq:discrete_basic_2}
\end{equation}
In RI-OAT, if the number of tomographic views is $J$ and the number of elements in the acoustic probe is $K_{trans}$.
The total number of transducer locations $Q = JK_{trans}$. 
The dimension of the system matrix in Eqn. (\ref{eq:discrete_basic_2}) is $M \times N = QL \times N = J K_{trans} L \times N$. 
The matrix $\mathbf{H}$ can be expressed as
\begin{equation}
	\mathbf{H} = \left[ \begin{array}{cccc} \mathbf{H}_1^T &\mathbf{H}_2^T &\ldots	&\mathbf{H}_J^T	 \end{array} \right] ^T ,
\end{equation}
where $\mathbf{H}_j \in \mathbb{R}^{K_{trans}L \times N}$ denotes the sub-matrix of $\mathbf{H}$ that maps the optical absorption map to the recorded data when the transducer is at the $j$-th view angle. 
We denote the discrete optical absorption energy density map at the $j$-th view angle as $\boldsymbol{\theta}_j$, and the pressure data recorded by the transducer array at this view angle as $\tilde{\mathbf{u}}_j$. 
The D-D RI-OAT imaging model is given by
\begin{align}
\mathbf{\tilde{u}} &= \left[ \begin{array}{c} \mathbf{\tilde{u}}_1  \\ \mathbf{\tilde{u}}_2 \\ \vdots \\ \mathbf{\tilde{u}}_J \end{array} \right]				
	= \left[ \begin{array}{cc} \mathbf{H}_1 &\boldsymbol{\theta}_1	\\ \mathbf{H}_2	&\boldsymbol{\theta}_2	\\	\vdots	&\vdots	\\ \mathbf{H}_J &\boldsymbol{\theta}_J	\end{array} \right]							\\ 				
																				 &= \left[ \begin{array}{cccc} \mathbf{H}_1 &\quad	&\quad &\quad	\\ \quad &\mathbf{H}_2 &\quad &\quad	 \\  \quad &\quad &\ddots &\quad  \\ \quad &\quad &\quad &\mathbf{H}_J	 \end{array}		\right] 
\left[ \begin{array}{cccc} \boldsymbol{\theta}_1	\\ \boldsymbol{\theta}_2 	\\ 	\vdots 	\\ \boldsymbol{\theta}_J	\end{array}	\right]
	\label{eq:inconsist_1}	\\
				&= \mathbf{H}_{aug} \boldsymbol{\Theta}, 		\label{eq:inconsist}
\end{align}
where $\boldsymbol{\Theta} = \left[ \begin{array} {ccc} \boldsymbol{\theta}^T_1, &\ldots, &\boldsymbol{\theta}^T_J \end{array} \right]^T $ is an augmented solution vector and $\mathbf{H}_{aug} \in \mathbb{R}^{M \times NJ}$ is an augmented system matrix that is  block diagonal and defined in terms of the view-specific system sub-matrices $\mathbf{H}_j$, $j=1,\cdots,J$.
 
Note that (\ref{eq:inconsist}) describes a mapping to the measurement data from a collection of $J$ estimates of object function, and these $J$ vectors will generally be distinct in RI-OAT. 
In current implementation of OAT reconstruction methods, it is generally desired to produce a single estimate of the object function that can be readily interpreted. 
And it is impractical to invert the D-D imaging equation (\ref{eq:inconsist}) and reconstruct $\boldsymbol{\Theta}$ for the following reasons. 
\begin{enumerate}
	\item First, the RI-OAT system matrix $\mathbf{H}_{aug}$ is highly rank deficient, making the problem ill-posed. 
	\item Second, the size of the augmented system matrix $\mathbf{H}_{aug}$, which is $J$ times larger than the original $\mathbf{H}$ in conventional OAT, adds an extremely heavy computational burden to any iterative reconstruction method. 
\end{enumerate}
In practice and in this study, a conventional iterative reconstruction method based on Eqn. (\ref{eq:discrete_basic}) is employed. 

\subsubsection{Interpretation of image reconstructed  using a discrete form of the FBP method}
The discretized form of the FBP method used in Section \ref{subsec:image_recon_method} can be decomposed into the product of two operators: the adjoint operator of $\mathbf{H}$ and a filtering matrix $\mathbf{F}$. 
Here, we assume that the filter independently acts on the measurement data corresponding to each transducer element.
The estimated solution can be expressed as 

\begin{gather}
	\hat{\boldsymbol{\theta}}^{RI} = \mathbf{H}^\dagger \mathbf{F} \mathbf{\tilde{u}}_{meas}		\\
= \mathbf{H}^\dagger \mathbf{F}  \left[ \begin{array}{cccc}	\mathbf{\tilde{u}}_1^T 	&\mathbf{\tilde{u}}_2^T &\ldots &\mathbf{\tilde{u}}_J^T	\end{array}	\right] ^T
		=	\mathbf{H}^\dagger \left[ \begin{array}{cccc}	\mathbf{\tilde{v}}_1^T 	&\mathbf{\tilde{v}}_2^T &\ldots &\mathbf{\tilde{v}}_J^T	\end{array}	\right] ^T, 
	\label{eq:append_basic}
\end{gather}
where $\mathbf{H}^\dagger$ denotes the adjoint operator of $\mathbf{H}$, $\mathbf{\tilde{u}}_j$ is the measurement data recorded by the transducer at $i$-th view angle, and $\mathbf{\tilde{v}}_j$ is $\mathbf{\tilde{u}}_j$ filtered by $\mathbf{F}$.  
Define the masking matrix $\mathbf{M}_k$ as 
\begin{equation}
\mathbf{M}_k = \left[ \begin{array} {ccccccc} \mathbf{0}_{L \times L}, &\ldots, &\mathbf{0}_{L \times L}, &\mathbf{I}_{L \times L}, &\mathbf{0}_{L \times L}, &\ldots, &\mathbf{0}_{L \times L} \end{array} \right] 		\label{eq:mask}
\end{equation}
where there are $(k-1)$ $\mathbf{0}_{L \times L}$ matrices before the identity matrix $\mathbf{I}_{L \times L}$, and $(J-k)$ $\mathbf{0}_{L \times L}$ matrices after it. 
By applying the adjoint operator of the augmented system matrix $\mathbf{H}_{aug}$ to the measurement data, one obtains
\begin{gather}
	\boldsymbol{\Theta}_{est} = \mathbf{H}_{aug}^\dagger \mathbf{F} \tilde{\mathbf{u}}_{meas}			\\
= \left[ \begin{array}{c c c c}	\mathbf{H}^\dagger &\quad &\quad &\quad \\		\quad &\mathbf{H}^\dagger &\quad &\quad \\		\quad &\quad &\ddots &\quad	\\		\quad &\quad &\quad &\mathbf{H}^\dagger		\end{array}	\right]				
\left[	\begin{array}{cccc}	\mathbf{M}_1^\dagger &\quad &\quad &\quad \\		\quad &\mathbf{M}_2^\dagger	&\quad	&\quad	\\		\quad &\quad &\ddots &\quad	\\	\quad &\quad &\quad	&\mathbf{M}_J^\dagger		\end{array}	\right]		
	\mathbf{F}
\left[	\begin{array}{c}	\tilde{\mathbf{u}}_1 \\ \tilde{\mathbf{u}}_2 	\\ \vdots \\	\tilde{\mathbf{u}}_J		\end{array}	\right]			\\
= \left[ \begin{array}{c c c c}	\mathbf{H}^\dagger &\quad &\quad &\quad \\		\quad &\mathbf{H}^\dagger &\quad &\quad \\		\quad &\quad &\ddots &\quad	\\		\quad &\quad &\quad &\mathbf{H}^\dagger		\end{array}	\right]				
\left[	\begin{array}{cccc}	\mathbf{M}_1^\dagger &\quad &\quad &\quad \\		\quad &\mathbf{M}_2^\dagger	&\quad	&\quad	\\		\quad &\quad &\ddots &\quad	\\	\quad &\quad &\quad	&\mathbf{M}_J^\dagger		\end{array}	\right]		
\left[	\begin{array}{c}	\tilde{\mathbf{v}}_1 \\ \tilde{\mathbf{v}}_2 	\\ \vdots \\	\tilde{\mathbf{v}}_J		\end{array}	\right]			\\
=\left[	\begin{array}{cccc} (\mathbf{H}^\dagger \left[ \begin{array}{c} \mathbf{\tilde{v}}_1	\\	\mathbf{0} \\ \vdots \\ \mathbf{0}	\end{array}	\right])^T 
	&(\mathbf{H}^\dagger \left[ \begin{array}{c} \mathbf{0}	\\	\mathbf{\tilde{v}}_2 \\ \vdots \\ \mathbf{0}	\end{array}	\right])^T
	&\ldots		
	&(\mathbf{H}^\dagger \left[ \begin{array}{c} \mathbf{0}	\\	\mathbf{0} \\ \vdots \\ \mathbf{\tilde{v}}_J	\end{array}	\right]) ^T
		\end{array}	\right]^T			\\
	= \left[	\begin{array}{cccc} \boldsymbol{\hat{\theta}}_1^T 	&\boldsymbol{\hat{\theta}}_2^T	&\ldots	 &\boldsymbol{\hat{\theta}}_J^T	\end{array}	\right]^T	.			\label{eq:append_1}
\end{gather}
Then, by comparing Eqn. (\ref{eq:append_basic}) and Eqn. (\ref{eq:append_1}), one obtains
\begin{gather}
	\hat{\boldsymbol{\theta}}^{RI} = \sum_{j=1}^{J} {\boldsymbol{\hat{\theta}}}_j,  
	\label{eq:append_2}
\end{gather}
Eqn. (\ref{eq:append_2}) is the discrete analog of Eqn. (\ref{eq:RI_summation}).
This indicates that when the discrete form of the FBP method is employed, the image reconstructed in RI-OAT, $\hat{\boldsymbol{\theta}}^{RI}$, can be interpreted as a superposition of estimates of $\hat{\boldsymbol{\theta}}$ that are reconstructed from each of the $J$ limited-view subproblems. 

\subsection{Optical properties used in simulation}
\label{append:OpticalProperty}
In Table \ref{tab:optical_properties} we list the optical properties we used in the Monte Carlo simulations.

\begin{table}[htbp]
	\centering
	\begin{threeparttable}[htbp]
		\caption{Representative Optical Properties Used in Monte Carlo Simulation}
		\label{tab:optical_properties}
		\begin{tabular}{l | l | l | l | l}
			\hline
			Tissue type  	&$\mu_a$(cm$^{-1}$) \tnote{1} 	&$\mu_s$(cm$^{-1}$) \tnote{2}		&$g$\tnote{3}		&$n$\tnote{4}		\TStrut \BStrut	\\	\hline
			Water						&0		  &0										&0.99		&1.33		\TStrut	\\
			Normal breast 	&0.03 	&80										&0.95		&1.40		\\
			Breast tumor		&0.3 		&100										&0.95		&1.40		\\
			Blood vessel		&9 			&179										&0.95		&1.40		\\
			Skin						&0.08		&93											&0.80		&1.40		\BStrut	\\	\hline
		\end{tabular}

		\begin{tablenotes}
		\item[1] $\mu_a$: optical absorption coefficient. 
		\item[2] $\mu_s$: optical scattering coefficient. 
		\item[3] $g$: scattering anisotropy. 
		\item[4] $n$: refractive index. 
		\end{tablenotes}

	\end{threeparttable}
\end{table}

\bibliography{ref}
\bibliographystyle{abbrv}





\end{document}